\documentclass[draft,jgrga]{AGUTeX}
\usepackage{graphicx}
\usepackage{amsfonts}
\usepackage{color}
\usepackage{lineno}
\linenumbers*[1]

\authorrunninghead{XIONG ET AL.}
\titlerunninghead{MC-MC INTERACTION AND ITS GEOEFFECTIVENESS}

\authoraddr{Ming Xiong, Huinan Zheng, Yuming Wang, and Shui Wang,
School of Earth and Space Sciences, University of Science and
Technology of China, Hefei, Anhui 230026, China.
(mxiong@ustc.edu.cn; hue@ustc.edu.cn; ymwang@ustc.edu.cn; and
swan@ustc.edu.cn)}

\authoraddr{S. T. Wu,
Center for Space Plasma and Aeronomic Research, University of Alabama in Huntsville, Huntsville,
Alabama 35899, USA. (wus@cspar.uah.edu)}

\begin{document}
\setkeys{Gin}{draft=false}

\title{Magnetohydrodynamic simulation of the
interaction between two interplanetary magnetic clouds and its
consequent geoeffectiveness}



\author{Ming Xiong, \altaffilmark{1,2} Huinan Zheng, \altaffilmark{1,2} S. T. Wu, \altaffilmark{3}
Yuming Wang, \altaffilmark{1} and Shui Wang\altaffilmark{1}}

\altaffiltext{1}
{Chinese Academy of Sciences Key Laboratory for Basic Plasma Physics, School of Earth
and Space Sciences, University of Science and Technology of China, Hefei, Anhui 230026, China}

\altaffiltext{2}{State Key Laboratory of Space Weather, Chinese Academy of Sciences, Beijing 100080,
China}

\altaffiltext{3}{Center for Space Plasma and Aeronomic Research, University of Alabama in Huntsville,
Huntsville, Alabama 35899, USA}

\begin{abstract}
Numerical studies of the interplanetary ``multiple magnetic clouds
(Multi-MC)" are performed by a 2.5-dimensional ideal
magnetohydrodynamic (MHD) model in the heliospheric meridional
plane. Both slow MC1 and fast MC2 are initially emerged along the
heliospheric equator, one after another with different time
interval. The coupling of two MCs could be considered as the
comprehensive interaction between two systems, each comprising of
an MC body and its driven shock. The MC2-driven shock and MC2 body
are successively involved into interaction with MC1 body. The
momentum is transferred from MC2 to MC1. After the passage of
MC2-driven shock front, magnetic field lines in MC1 medium
previously compressed by MC2-driven shock are prevented from being
restored by the MC2 body pushing. MC1 body undergoes the most
violent compression from the ambient solar wind ahead, continuous
penetration of MC2-driven shock through MC1 body, and persistent
pushing of MC2 body at MC1 tail boundary. As the evolution
proceeds, the MC1 body suffers from larger and larger compression,
and its original vulnerable magnetic elasticity becomes stiffer
and stiffer. So there exists a maximum compressibility of Multi-MC
when the accumulated elasticity can balance the external
compression. This cutoff limit of compressibility mainly decides
the maximally available geoeffectiveness of Multi-MC, because the
geoeffectiveness enhancement of MCs interacting is ascribed to the
compression. Particularly, the greatest geoeffectiveness is
excited among all combinations of each MC helicity, if magnetic
field lines in the interacting region of Multi-MC are all
southward. Multi-MC completes its final evolutionary stage when
the MC2-driven shock is merged with MC1-driven shock into a
stronger compound shock. With respect to Multi-MC
geoeffectiveness, the evolution stage is a dominant factor,
whereas the collision intensity is a subordinate one. The magnetic
elasticity, magnetic helicity of each MC, and compression between
each other are the key physical factors for the formation,
propagation, evolution, and resulting geoeffectiveness of
interplanetary Multi-MC.
\end{abstract}
\begin{article}


\section{Introduction}
Space weather refers to the conditions on the Sun and in the solar
wind, magnetosphere, ionosphere, and thermosphere that can
influence the performance and reliability of space-borne and
ground-based technological systems or can endanger human life or
health, as defined in US National Space Weather Program
Implementation Plan. A seamless forecasting system for Space
weather lies on the comprehensive and in-depth understanding of
the Sun-Earth system. The never-stopping tremendous efforts have
been made by humankind since the space age of the 1950s. A great
deal of the sophisticated observations beyond the Earth are now
provided, with the launching of various spacecraft into deep
space, such as Yohkoh, Geotail, Wind, SOHO, Ulysses, ACE, TRACE in
the 1990s, and Cluster, RHESSI, SMEI, DS, Hinode (Solar B), STEREO
in the 21st century. These spacecraft missions construct an
indispensable backbone in the establishment of space weather
prediction system. Meanwhile, many models have been or are being
developed and applied to space weather forecasting by utilizing
most measurements of the above spacecraft, such as (1) HAF
(Hakamada-Akasofu-Fry) \citep{Fry2001,Fry2005}; (2) STOA (Shock
Time of Arrival) \citep{Smart1985}; (3) ISPM (Interplanetary Shock
Propagation Model) \citep{Smith1990}; (4) an ensemble of HAF, STOA
and ISPM models \citep{Dryer2001,Dryer2004,McKenna-Lawlor2006};
(5) SPM (Shock Propagation Model) \citep{Feng2006}; (6) SWMF
(Space Weather Modeling Framework) \citep{Toth2005}; (7) HHMS
(Hybrid Heliospheric Modeling System) \citep{Detman2006}; (8) a
data-driven Magnetohydrodynamic (MHD) model of the University of
Alabama in Huntsville \citep{Wu2005a,Wu2006a}; (9) a 3D regional
combination MHD model with inputs of the source surface
self-consistent structure based on the observations of the solar
magnetic field and K-coronal brightness \citep{Shen2007}; (10) A
merging model of SAIC MAS and ENLIL Heliospheric MHD Model
\citep{Odstrcil2004b}; (11) an HAF + 3-D MHD model
\citep{Wu2005c,Wu2006c,Wu2007b,Wu2007c}, and so on. However, great
challenges are still faced to improve the prediction performance
of space weather, as human civilization is relying more and more
on space environment \citep{Baker2002,Fisher2004}.

The interplanetary (IP) space is a pivot node of the
solar-terrestrial transport chain. Solar transients, e.g., shocks
and coronal mass ejections (CMEs), propagate in it, interact with
it, and cause many consequences in the geo-space. Magnetic clouds
(MCs) are an important subset of interplanetary CMEs (ICMEs),
occupying the fraction of nearly $\sim$ 100\% (though with low
statistics) at solar minimum and $\sim$ 15\% at solar maximum
\citep{Richardson2004,Richardson2005}, and have significant
geoeffectiveness
\citep{Tsurutani1988,Gosling1991,Gonzalez1999,Wu2002a,Wu2002b,Wu2003,Wu2006b,Huttunen2005}.
The current intense study of MCs could be traced back to the
pioneer work by \citet{Burlaga1981}, who firstly defined an MC
with three distinct characteristics of enhanced magnetic field
strength, smooth rotation of magnetic field vector, and low proton
temperature, and described it as a flux rope structure. An MC is
widely thought to be the IP manifestation of a magnetic flux rope
in the solar corona, which loses equilibrium and then escapes from
the solar atmosphere into the IP space \citep{Forbes2006}, with
its both ends still connecting to the solar surface
\citep{Larson1997}.

It is very likely for solar transients to interact with each other
on their way to the Earth, especially at solar maximum when the
daily occurrence rate of CMEs is about 4.3 in average on basis of
the SOHO/Lasco CME catalogue
(http://cdaw.gsfc.nasa.gov/CME\_list). Some IP complicated
structures were reported, such as complex ejecta
\citep{Burlaga2002}, multiple MCs (Multi-MC)
\citep{Wang2002,Wang2003a}, shock-penetrated MCs
\citep{Wang2003b,Berdichevsky2005,Collier2007},
non-pressure-balanced ``MC boundary layer'' associated with
magnetic reconnection \citep{Wei2003a,Wei2003b,Wei2006}, ICMEs
compressed by a following high-speed stream \citep{DalLago2006},
multiple shock interactions \citep{Wu2005d,Wu2006d,Wu2007a}.
However, all space-borne instruments, except the heliospheric
imagers onboard SMEI and STEREO, observe either the solar
atmosphere within 30 solar radii by remote sensing, or the in-situ
space by local detecting, or both. Thus, numerical simulations are
necessary to understand the whole IP dynamics. Below is an
incomplete list of numerical studies of dynamical processes of
CMEs/MCs and complex structures in the IP medium mentioned before:
an individual CME/MC
\citep{Vandas1995,Vandas1996,Vandas2002,Groth2000,
Schmidt2003,Odstrcil2003,Odstrcil2004a,Odstrcil2005,Manchester2004,Wu2005b},
the interaction of a shock wave with an MC
\citep{Vandas1997,Xiong2006a,Xiong2006b}, the interaction of
multiple shocks \citep{Wu2004a,Wu2004b,Wu2005d,Wu2006d,Wu2007a},
and the interaction of multiple ejecta
\citep{Gonzalez-Esparza2004,Gonzalez-Esparza2005,Lugaz2005,
Xiong2005,Wang2004,Wang2005b,Wu2006c,Wu2007c,Hayashi2006}.
Therein, \citet{Wu2005d,Wu2006d,Wu2007a} performed a 1.5-D MHD
model to simulate the famous Halloween 2003 epoch, in which
eruption time of solar flares was used as input timing for solar
disturbances to study the shock-shock interaction (and overtaking)
and the matching of shock arrival time at 1 AU with observations
(ACE). In addition, \citet{Wu2006c,Wu2007c} performed 3-D global
simulations by combining two simulation models (HAF + 3-D MHD) to
study the interacting and overtaking of two ICMEs. These
observation and simulation efforts do advance our understanding of
solar-terrestrial physics.

The Multi-MCs have already been verified by observations to be an
important IP origin for the great geomagnetic storms
\citep{Wang2002,Wang2003a,Xue2005,Farrugia2006,Xie2006,Zhang2007}.
Particularly, for the 8 extremely large geomagnetic storms with
$Dst \leq$ $- 200$ nT during the year 2000 $\sim$ 2001, 2 of them
were caused by Multi-MCs and one caused by shock-MC interacting
structure \citep{Xue2005}. Most recently, via summarizing the
efforts of the NASA Living With a Star (LWS) Coordinated Data
Analysis Workshop (CDAW) held at George Mason University, in March
2005, \citet{Zhang2007} proposed that 24 out of 88 (27\%) major
geomagnetic storms with $Dst \leq $ $- 100$ nT from the year 1996
to 2005 were produced by multiple interacting ICMEs arising from
multiple halo CMEs launched from the Sun in a short period. So the
Multi-MC plays a notable role in producing large geomagnetic
storms. There are two possible conditions for double-MC formation
\citep{Wang2004}: (1) The speed of following MC should be faster
than that of preceding MC; (2) The separation between the eruption
of two MCs should be moderate (about 12 hours based on statistics
of observed events). Evolutionary signatures of ICMEs interacting
are found from spacecraft observations, i.e., heating of the
plasma, acceleration/deceleration of the leading/trailing ejecta,
compressed field and plasma in the leading ejecta, possible
disappearance of shocks, and strengthening of the shock driven by
the accelerated ejecta \citep{Farrugia2004}. Previous simulations
of interaction between two magnetic flux ropes in the IP space
\citep{Lugaz2005,Wang2005b}, the solar corona
\citep{Schmidt2004,Wang2005a,Lugaz2007}, and a local homogeneous
medium background \citep{Odstrcil2003} only address a few typical
cases in the dynamical aspect. Here a comprehensive study of many
cases of MCs interacting under various conditions is carried out
for better understanding of both dynamics and ensuing
geoeffectiveness. The interaction between two systems, each
comprising of an MC and its driven shock, could be considered in
some senses as a generalization of our recent studies of MC-shock
interaction \citep{Xiong2006a,Xiong2006b}. Thus we address the
following two issues naturally: (1) What is the role of the
following MC body in Multi-MC evolution in comparison with our
previous studies \citep{Xiong2006a,Xiong2006b} of MC-shock
interaction? (2) At what evolutionary stage a Multi-MC at 1 AU
reaches the maximum geoeffectiveness? The above answers are
explored by a 2.5-D numerical model in ideal MHD process.

The force-free magnetic flux rope models have been proven to be
very valuable to interpret in-situ observations of MCs
\citep[e.g.,][]{Lundquist1950,Burlaga1988,Farrugia1993,Chen1996,Owens2006}.
Particularly, Lundquist model \citep{Lundquist1950} is adopted in
our model to describe the magnetic field configuration of an MC,
as widely applied in the space science literature
\citep[e.g.,][]{Vandas1995,Vandas1996,Wang2002,
Wang2003d,Wang2005b,Xiong2006a,Xiong2006b}. A following fast MC
overtaking and interacting a preceding slow one in the IP space
could result in a Multi-MC structure \citep{Wang2002,Wang2003a}.
In order to explore the basic physics process of Multi-MC, we make
the following assumptions to simplify the complex circumstance of
double-MC structure in the numerical MHD simulation: (1) two MCs'
axes parallel or anti-parallel with each other; (2) their axes are
both within the ecliptic plane and perpendicular to the Sun-Earth
line; (3) each MC is symmetric in the azimuth direction of the
heliosphere, and considered as an ideal loop encompassing the Sun;
(4) magnetic reconnection does not exist in double MC interacting;
(5) both MCs have the same size, mass, magnetic field strength,
and plasma $\beta$. Thus, two MCs in our model only differ in
magnetic helicity sign $H_{mc}$ and initial radial lift-off speed
$v_{mc}$. A parametric study of $H_{mc}$ and $v_{mc}$ is focused
in our model for the very specialized Multi-MC structure. Since
the two MCs are very alike except $H_{mc}$ and $v_{mc}$, they
could be, to some extent, considered to be identical. MC1 and MC2
are respectively used to label the two MCs launched from the Sun,
one after another. Because an MC boundary is a self-enclosed
magnetic surface, and two MCs' magnetic field lines would not
blend under the condition of the strictly ideal MHD process, the
sub-structures of double MCs, corresponding to the previously
separated MC1 and MC2 before collision, could be easily
differentiated, and accordingly named as sub-MC1 and sub-MC2.

The goal of the present work is to conduct a systematic
investigation of Multi-MC in the IP space. We give a brief
description of the numerical MHD model in Section
\ref{Sec:Method}, describe the dynamical behavior of MC-MC
interaction in Section \ref{Sec:MC-MC}, discuss the consequent
geoeffectiveness in Section \ref{Sec:Geoeffect}, analyze the
compressibility of MC-MC collision in Section
\ref{Sec:Compressibility}, and summarize the paper in Section
\ref{Sec:Conclusion}.

%
%
\section{Numerical MHD Model}\label{Sec:Method}
The Multi-MC simulation is accommodated by a few slight
modifications from our previous numerical model for MC-shock
interaction \citep{Xiong2006a,Xiong2006b}. These modifications are
as follows, (1) The top boundary of simulated domain is extended
from 300 to 400 $R_s$; (2) The following shock is replaced by a
following MC; (3) The initial speed $v_{mc}$, emergence time
$t_{mc}$, and magnetic helicity $H_{mc}$ out of all input
parameters for each sub-MC of Multi-MC are independently selected
to make various combinations for parametric study shown in Table
\ref{Tab2}. First, the propagation through the IP space is modeled
by numerical simulation. Then, the geomagnetic storm excited by
the solar wind-magnetosphere-ionosphere coupling is approximated
by an empirical formula of Burton $\frac{dDst(t)}{dt} =Q(t) -
\frac{Dst(t)}{\tau}$ \citep{Burton1975}. Here the coupling
function $Q = v_r \cdot ~\mbox{Min}(B_z,0)$ and the diffusion time
scale $\tau=$ 8 hours, with the radial solar wind speed and
south-north magnetic field component respectively denoted by $v_r$
and $B_z$. Burton model \citep{Burton1975} for geomagnetic
disturbance has been analyzed and validated
\citep{Wang2003d,Wang2003e}, and applied in $Dst$ evaluation
\citep{Wang2003d,Wang2003e,Xiong2006a,Xiong2006b}. Thus the
physical process of cause-effect transport chain for solar
disturbances is fully described in our model. Moreover, the
MC2-driven shock in all of our simulation cases is faster than the
local magnetosonic speed all the way, and strong enough so that it
would not be dissipated in the low $\beta$ MC1 medium
\citep{Xiong2006a,Xiong2006b}.

\section{MC1-MC2 Interaction}\label{Sec:MC-MC}
All 48 simulation cases of MC1-MC2 coupling are assorted into 4
groups in Table \ref{Tab2}, with 18 cases of individual MC in 2
groups of Table \ref{Tab1} for comparison. Here, IM, EID, CID
respectively stand for ``Indiviual MC", ``Eruption Interval
Dependence", ``Collision Intensity Dependence", with the
subscripts 1 and 2 denoting the sign of magnetic helicity. Case
C$_1$ is shared by Groups EID$_1$ and CID$_1$, and Case C$_2$ by
Groups EID$_2$ and CID$_2$. In our simulation, an MC with
southward/northward magnetic field in its rear half is defined to
have positive/negative helicity. Both MCs are associated with
positive helicities in Groups EID$_1$ and CID$_1$, meanwhile MC1
and MC2 are respectively associated with positive and negative
helicities in Groups EID$_2$ and CID$_2$.

The numerical simulation is performed in the ideal MHD process.
The artificial numerical magnetic reconnection between MCs is
strictly ruled out by a specific numerical technique
\citep[c.f.][]{Hu2003b}. Thus the dynamics in Groups IM$_1$,
EID$_1$, and CID$_1$ is nearly the same as that in Groups IM$_2$,
EID$_2$, and CID$_2$ respectively, whereas the geoeffectiveness is
highly different due to the reversed north and south magnetic
components within the cloud with opposite helicity. Moreover, by
changing $Dt$ ($Dt=t_{mc2} - t_{mc1}$, $t_{mc1}=0$ hour), the
initiation delay between a preceding MC of 400 km/s and a
following MC of 600 km/s in Groups EID$_1$ and EID$_2$, the
Multi-MC formed by the MC1 and MC2 may reach different
evolutionary stages on its arrival at 1 AU. Therefore the eruption
interval dependence for MC1-MC2 interaction is easily
discriminated by a comparative study. Similarly, collision
intensity dependence is also explored by a parametric study of
$v_{mc2}$ from 450 to 1200 km/s in Groups CID$_1$ and CID$_2$.
Meanwhile the full interaction between sub-clouds within 1 AU to
maximally highlight collision effect is guaranteed by
$t_{mc2}=12.2$ hours in Groups CID$_1$ and CID$_2$. Furthermore,
an individual MC with its speed from 400 to 1200 km/s in Groups
IM$_1$ and IM$_2$ supplements indispensably to other Groups for
the study of coupling effect of two MCs. Cases B$_1$ and B$_2$
with $t_{mc2}=30.1$ hours, C$_1$ and C$_2$ with $t_{mc2}=12.2$
hours, are typical examples of Multi-MC in the early and late
evolutionary stages respectively, which are addressed below in
details.

\subsection{Case B$_1$}
\begin{figure}
\noindent
  \includegraphics[width=0.93\textheight,angle=90]{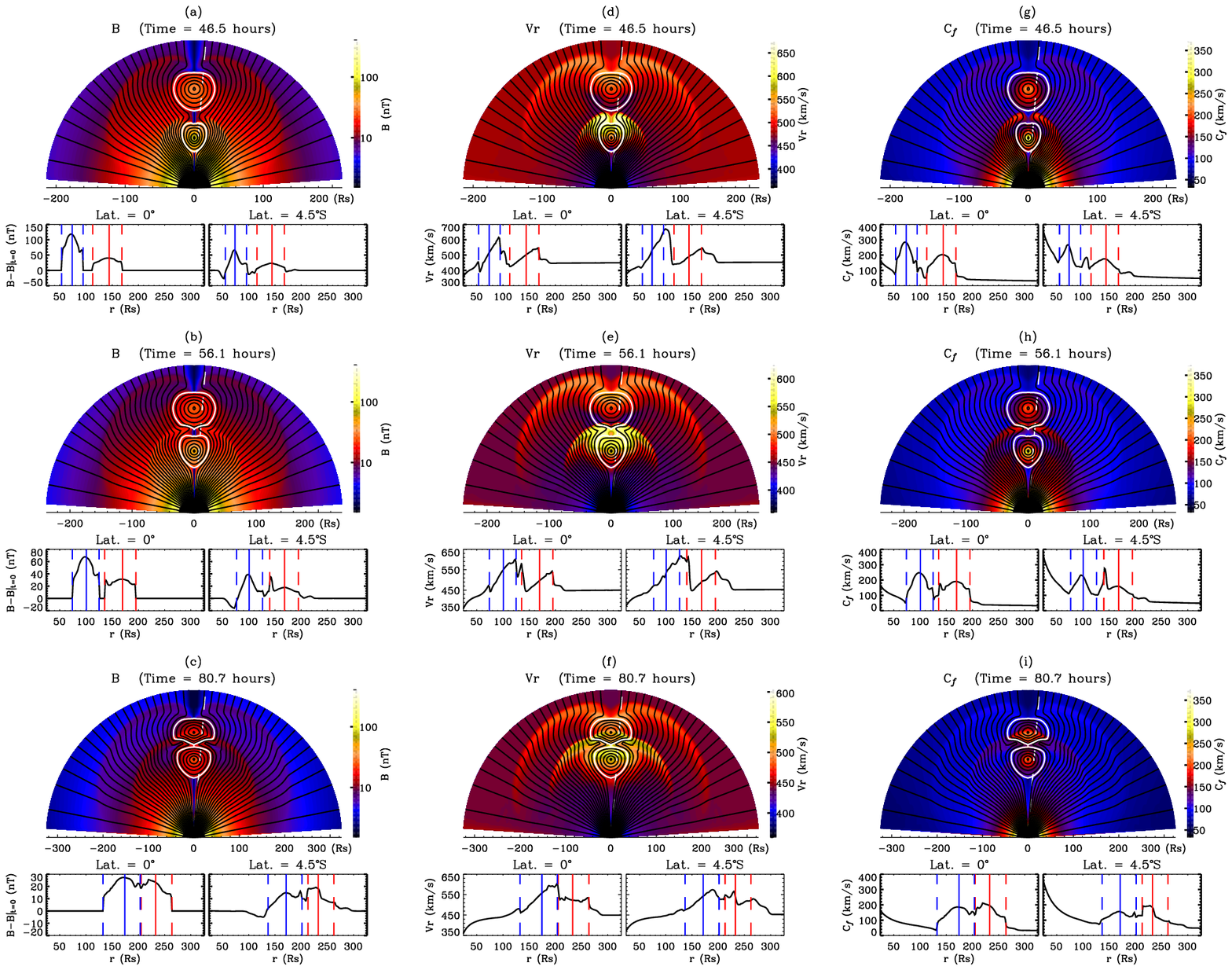}
\caption{} \label{case-B}
\end{figure}

In Case B$_1$, we discuss the results of MC1-MC2 interaction for
eruption speed $v_{mc1}=$ 400 km/s, $v_{mc2}=$ 600 km/s, and
initiation delay $t_{mc2}=$ 30.1 hours. Figure \ref{case-B} shows
the successive behavior of MC1-MC2 interaction of Case B$_1$. The
magnetic field lines, among which two are enclosed white solid
lines marking the boundaries of MC1 and MC2 respectively, are
superimposed on each color-filled contour image, and two radial
profiles, one through the equator (noted by Lat. = $0^\circ$), the
other through $4.5^\circ$ southward (white dashed lines in the
images, noted by Lat. = $4.5^\circ$S), are plotted below. One can
read the global vision from the images and local details from the
profiles simultaneously for the propagation and evolution of
Multi-MC. For better highlighting the local disturbance, Figures
\ref{case-B}(a)-(c) show the magnitude $B$ of magnetic field from
which the initial value $B|_{t=0}$ is deducted. Two identical MCs
are successively injected into the IP space with different initial
eruption speed. As long as the fast following MC2 lags behind the
slow preceding MC1, each of them behaves as an individual event,
and satisfies the criteria of a single MC. Because the MC-driven
shock and incidental shock \citep{Xiong2006a} both propagate along
the heliospheric current sheet (HCS) in the IP medium, their
inherent traits are identically characterized by a concave-outward
morphology with the position of the strongest intensity being
roughly $4.5^\circ$ away from the HCS. MC2-driven shock just
approaches MC1 body tail at 46.5 hours, as seen from Figure
\ref{case-B}(d). Across this shock front, radial speed $v_r$
increases abruptly from 440 km/s at MC1 tail to 670 km/s at MC2
head. From then on, MC2 and MC1 will directly collide to form a
special IP complex named Multi-MC by \citet{Wang2002,Wang2003a},
and their evolution will be coupled with each other. Consequently,
the characteristic parameters of each sub-MC would change
drastically due to the non-linear interaction. At 56.1 hours,
MC2-driven shock front has already entered MC1 body across which
radial speed $v_r$ abruptly jumps from 445 to 620 km/s, but MC2
body is still unable to catch up with MC1 tail (Figure
\ref{case-B}(e)) because of $t_{mc2}=$ 30.1 hours. The dynamic
response of Multi-MC at this snapshot is merely ascribed to the
interaction between MC2-driven shock and MC1 body. So the
preceding MC1 behavior in Figures \ref{case-B}(b), (e), and (h)
are similar to its counterpart of MC-shock interaction in essence
\cite[c.f. Figures 3(c), (f), and (i) in][]{Xiong2006a}. Large
compression within MC1 medium downstream of MC2-driven shock front
is very pronounced from an abnormal local spike-like structure of
$c_f$ along Lat. = $4.5^\circ$S, as shown in Figure
\ref{case-B}(h). The orientation of magnetic field lines is also
rotated in MC1 medium swept by the shock front. As the shock
continuously advances into MC1 body, the morphology of MC1 rear
part is transformed from an original rough semi-circle (Figure
\ref{case-B}(d)) to a V-shape with a wide open mouth (Figure
\ref{case-B}(f)). Moreover MC2 body has already contacted MC1 tail
at the bottom of so-called V shape along the equator at 80.7
hours, when the MC2-driven shock cannibalizes the rear half of MC1
body (Figure \ref{case-B}(f)). Since then, MC2 body is directly
involved into interaction with MC1 body. The Multi-MC evolution
has reached a new critical stage, for MC1 will undergoes the most
violent compression from the ambient solar wind ahead, continuous
penetration of MC2-driven shock through MC1 body, and persistent
pushing of MC2 body at MC1 tail boundary. In Figure
\ref{case-B}(f), nearly constant speed in MC1 rear half and large
speed difference with 80 km/s across MC1 rear boundary along the
equator imply continuous strike of high-speed MC2 body upon
preceding MC1 body. Besides, the interplanetary magnetic field
(IMF) within Multi-MC envelope is highly bending just behind
MC2-driven shock front (Figures \ref{case-B}(c), (f), and (i)), as
a result of rotation across the shock front and draping around
either sub-cloud surface.

\begin{figure}
\noindent
   \includegraphics[width=20pc]{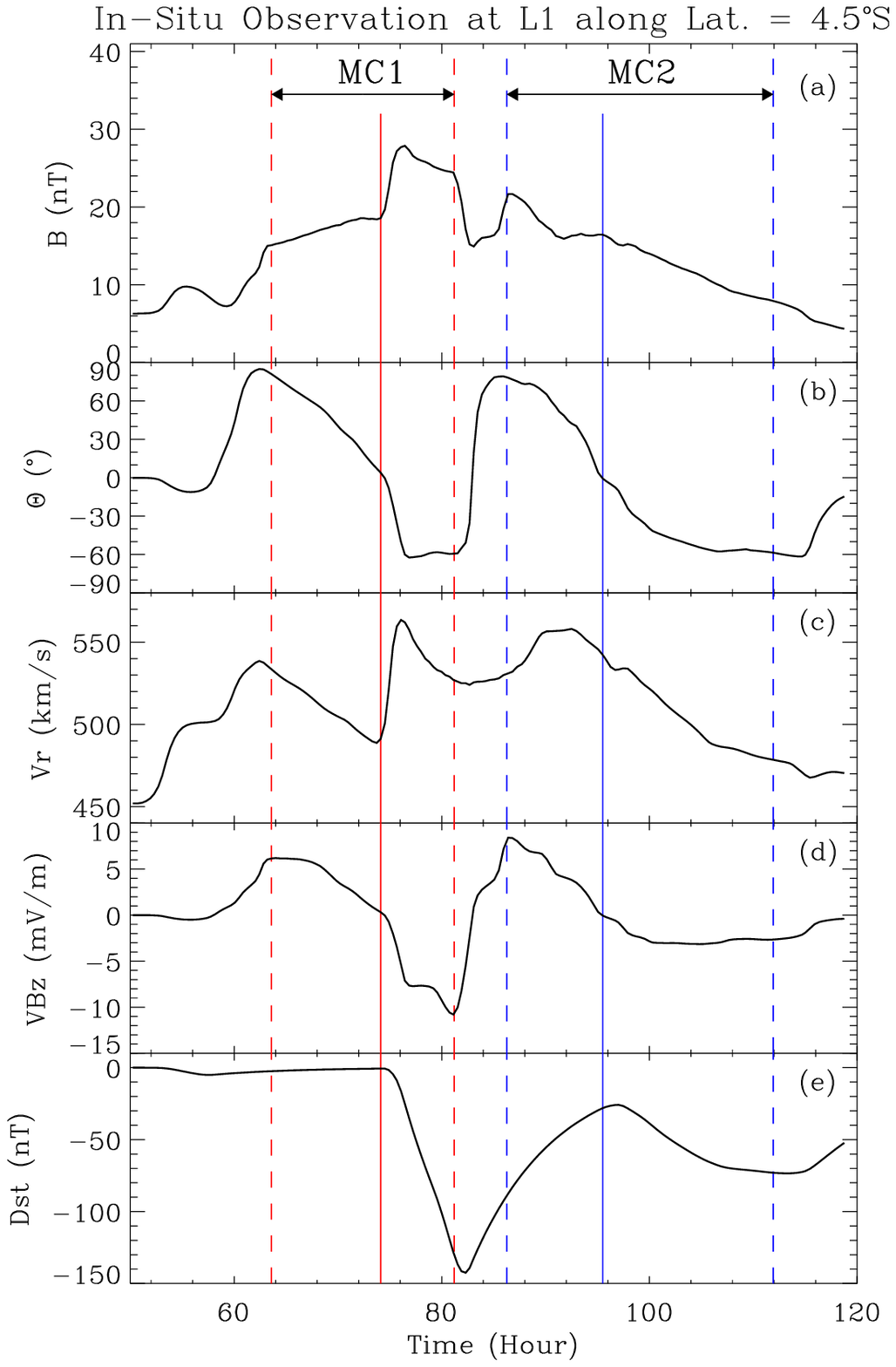}
\caption{} \label{Satellite-B}
\end{figure}

The in-situ observation along Lat. = $4.5^\circ$S by a
hypothetical spacecraft at Lagrangian point (L1) is illustrated in
Figure \ref{Satellite-B}. With each sub-MC boundary identified as
dashed lines, the MC1 duration of 18 hours is much less than MC2
duration of 26 hours due to the compression in MC1 rear half
accompanying with MC2-driven shock advancing. The MC2 ``senses''
the existence of preceding MC1, though its response is much less
sensitive. The location of maximum bulk flow speed $v_r$ in MC2
body is shifted by 6 hours later (Figure \ref{Satellite-B}(c))
from MC2 head \citep[c.f. Figure 2 in][]{Xiong2006a}, between
which magnitude $B$ is obviously enhanced (Figure
\ref{Satellite-B}(a)). The dawn-dusk electric field $VB_z$ is
calculated by the product of $v_r$ and $B_\theta$ in the spherical
geometry of this simulation. Beginning from 74 hours, $VB_z$,
negative in MC1 rear half, positive in MC2 front half, and
negative again in MC2 rear half (Figure \ref{Satellite-B}(d)), is
respectively responsible for $Dst$ dropping from 0 nT at 74 hours
to $-140$ nT at 82 hours, recovering from $-140$ nT at 82 hours to
$-25$ nT at 97 hours, and dropping again from $-25$ nT at 97 hours
to $-75$ nT at 114 hours (Figure \ref{Satellite-B}(e)). Owing to
compression of southward magnetic component $B_s$
($B_s=~\mbox{Min}(B_\theta,0)$) within MC1 rear part, the first
$Dst$ dip with $-140$ nT is much lower than the second one with
$-75$ nT for geomagnetic storm. Particularly, the two $Dst$ dips
are separated by only 32 hours, because the geoeffectiveness of
two IP triggers (MC1 and MC2) are superposed together. The idea of
a two-ejecta event associated with a two-step geomagnetic storm
was recently proposed and verified by \citet{Farrugia2006} on
basis of observation. Hence the association of two $Dst$ dips lies
in the MC1-MC2 interaction.

\subsection{Case C$_1$}\label{Sec:CaseC}
\begin{figure}
\noindent
    \includegraphics[width=20pc]{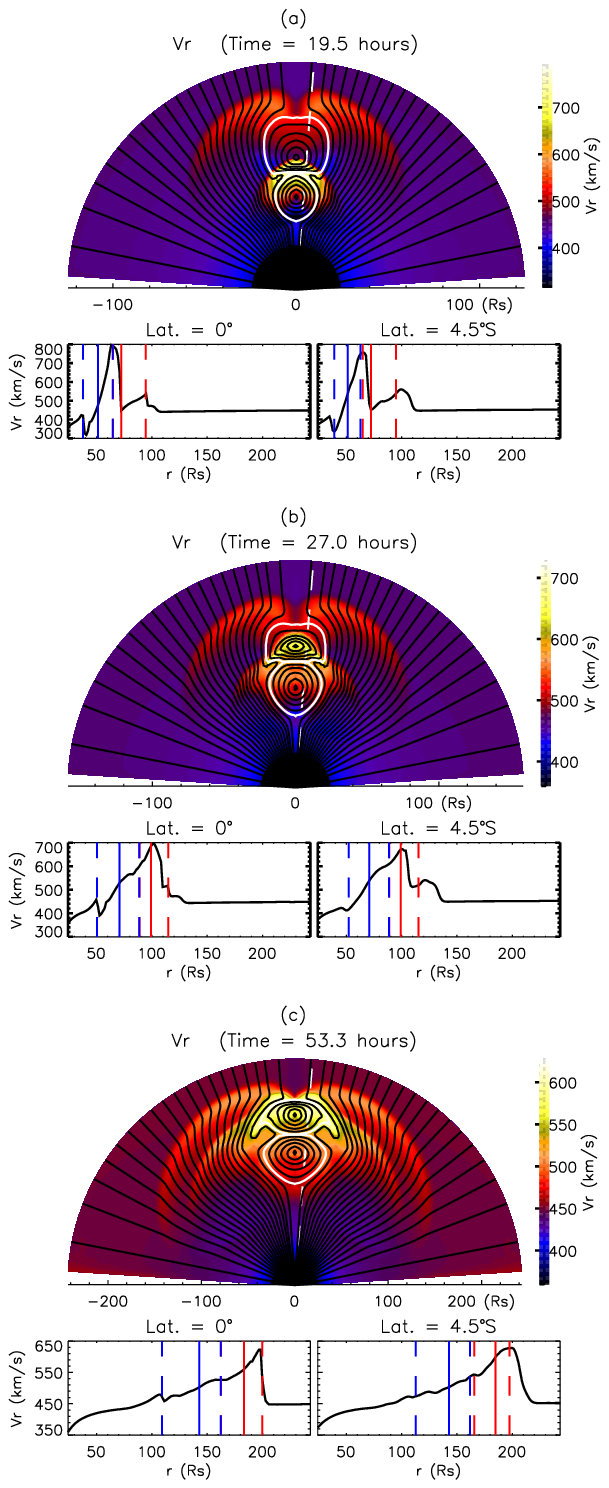}
\caption{} \label{Case-C}
\end{figure}

In order to realize the fully interaction between MC1 and MC2
before their arrival at L1, $t_{mc2}$, the emergence time of MC2,
is scheduled earlier to be 12.2 hours with both MCs having the
same speeds of Case B$_1$. Only the evolution of $v_r$ is given in
Figure \ref{Case-C} to visualize multi-cloud structure. Comparing
to that in Figures \ref{case-B}(c), (f), and (i), the so-called
``V-shape" morphology of MC1 rear half becomes very flat under the
pounding of very high-speed MC2 body at 19.5 hours as Multi-MC
evolution proceeds, as shown in Figure \ref{Case-C}(a). As a
result, contact position between MC1 and MC2 body is extended from
one single point at the HCS (Figure \ref{case-B}(f)) to a straight
line between Lat. = $4.5^\circ$S and $4.5^\circ$N (Figure
\ref{Case-C}(a)). The MC1's magnetic elasticity seems to be too
vulnerable to resist the violent collision from MC2 body. The
collision efficiently transfers the radial momentum from the fast
following MC2 to the slow preceding MC1. It results in
monotonically decreasing $v_r$ from the head to tail of Multi-MC
at 53.3 hours, resembling a single MC, as seen in Figure
\ref{Case-C}(c). Besides, MC2 morphology turns from a
radial-extent-elongated ellipse (Figure \ref{Case-C}(a)) to an
angular-extent-elongated one (Figure \ref{Case-C}(c)) due to the
blocking of MC1 body ahead. MC2 body is also compressed radially
to some extent. Certainly, the compression of MC2 body is much
less than that of MC1 body. Moreover, MC2-driven shock ultimately
penetrates the MC1 body (Figure \ref{Case-C}(c)), and will merge
with the MC1-driven shock into a stronger compound shock, which is
consistent with the previous results of double MC interaction
\citep{Odstrcil2003,Lugaz2005}. Therefore the Multi-MC has nearly
been completing its final evolutionary stage at 53.3 hours, after
which the Multi-MC will move forwards as a relatively stable
structure.

Time sequence of hypothetical measurement at L1 for Case $C_1$ is
shown in Figure \ref{Satellite-C}. The MC2-driven shock just
emerges from MC1 body after penetrating it, so no extremum of
speed profile $v_r$ is found inside the multi-cloud. Double dips
of $Dst$ index are $-93$ nT and $-95$ nT, increased by 47 nT and
decreased by 20 nT respectively in contrast with those in Case
B$_1$ in Figure \ref{Satellite-B}(e). The mitigation of
geoeffectiveness for the first $Dst$ dip is owing to the position
of MC2-driven shock front far away from the rear part of MC1 with
southward magnetic component, the aggravation for the second $Dst$
dip is ascribed to the MC2 body compression mentioned above. A
peak of $VB_z$ up to 14 mV/m can be seen near the MC1 front
boundary, where the largest compression occurs. However, it is
positive and makes no contribution to geoeffectiveness.
Additionally, the durations of MC1 and MC2 are shortened by 4.7
and 3 hours, respectively, as compared with those in Case B$_1$.

Figure \ref{multi-geometry} shows the time-dependent parameters of
Multi-MC Case C$_1$ (thick curves), where the dotted, dashed, and
dotted vertical lines from left to right denote the occasion of
MC2-driven shock encountering MC1 body tail, MC2 body hitting MC1
body tail, and MC2-driven shock reaching MC1 body head,
respectively. Two corresponding isolated MC cases are superimposed
as thin curves for comparison. The acceleration of MC1 is large
and early, while the deceleration of MC2 is small and late, as
seen from Figure \ref{multi-geometry}(a). The radial compression
of MC2 body brings not only the shortening of its radial span $Sr$
but also the stretching of its angular span $S\theta$. The
behavior of MC1 is a bit more complex. In our previous studies of
MC-shock interaction \citep{Xiong2006a,Xiong2006b}, MC compressed
morphology will be restored after the shock passage. However, in
the presence of the following MC2 body's pushing effect for
Multi-MC case, MC2 body will take over the role of suppressing
MC1's inherent magnetic elasticity when MC2-driven shock moves
farther. Thus the firm gripping of MC1 body at all time leads to
significant shrinking of its volume. It is why cross section area
of MC1 body is smaller than that of MC2 (Figure
\ref{multi-geometry}(d)). Particularly, as seen from the local
minimum value of $Sr$ at 30 hours in Figure
\ref{multi-geometry}(b), the compression of MC1's $Sr$ reaches to
its extreme when the MC2-driven shock nearly arrives at MC1 head
boundary. Meanwhile the temporarily enhanced $S\theta$ of MC1
during 24 $\sim$ 38 hours is steadily reduced afterwards (Figure
\ref{multi-geometry}(c)).

\subsection{MC2 Helicity Role}
\begin{figure}
\noindent
   \includegraphics[width=20pc]{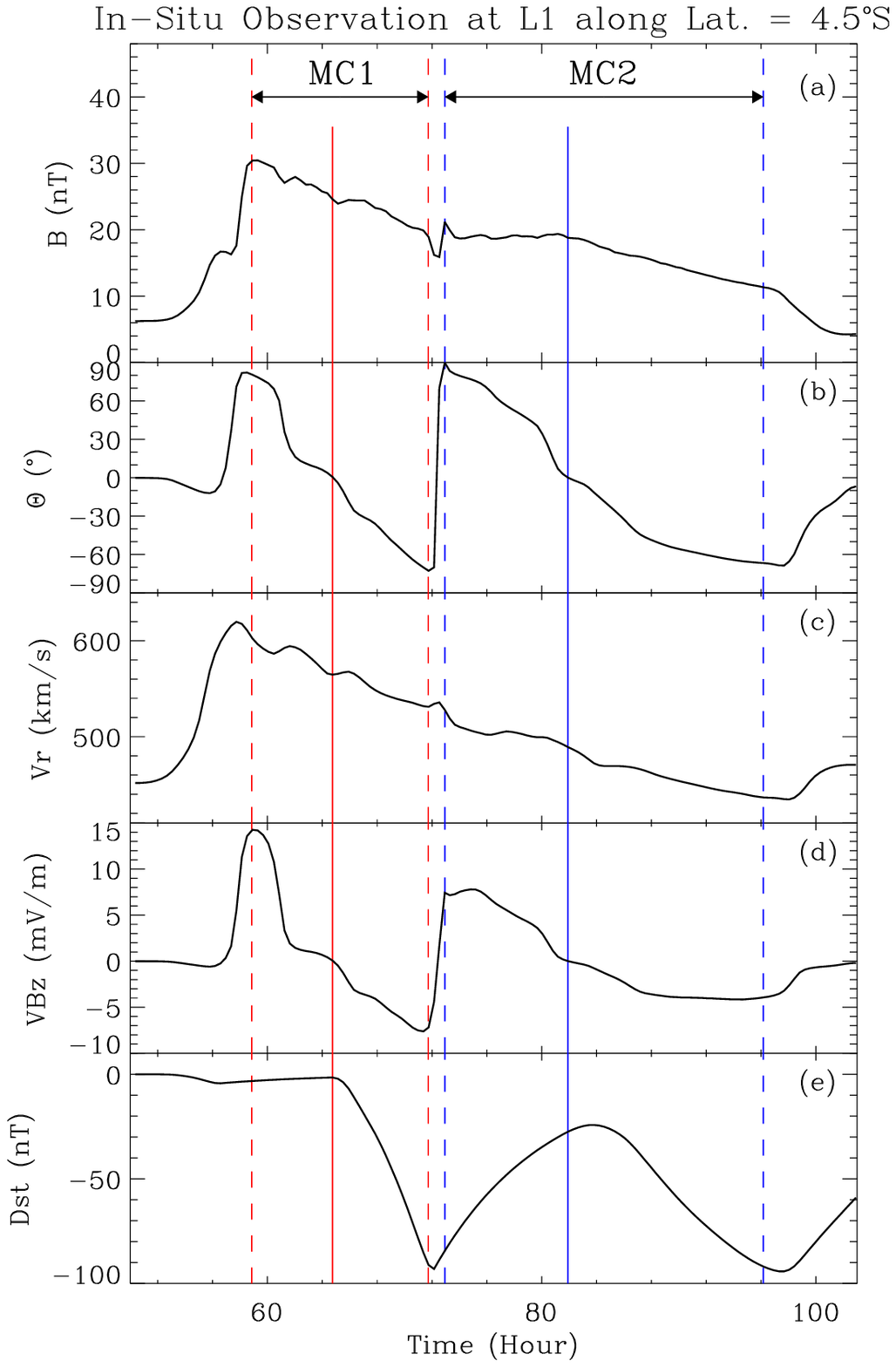}
\caption{} \label{Satellite-C}
\end{figure}

\begin{figure}
   \includegraphics[width=20pc]{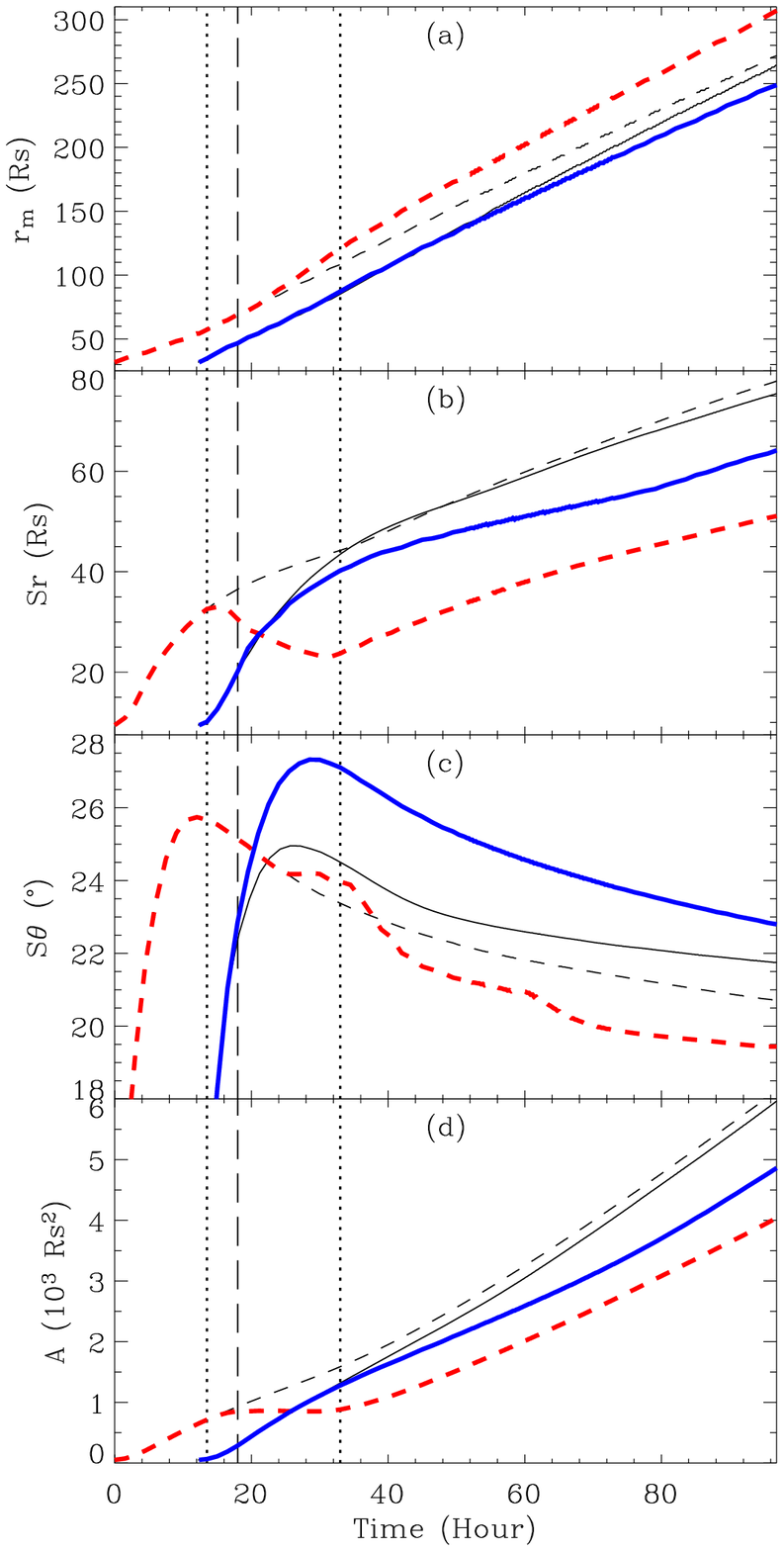}
\caption{} \label{multi-geometry}
\end{figure}

\begin{figure}
   \includegraphics[width=0.99\textwidth]{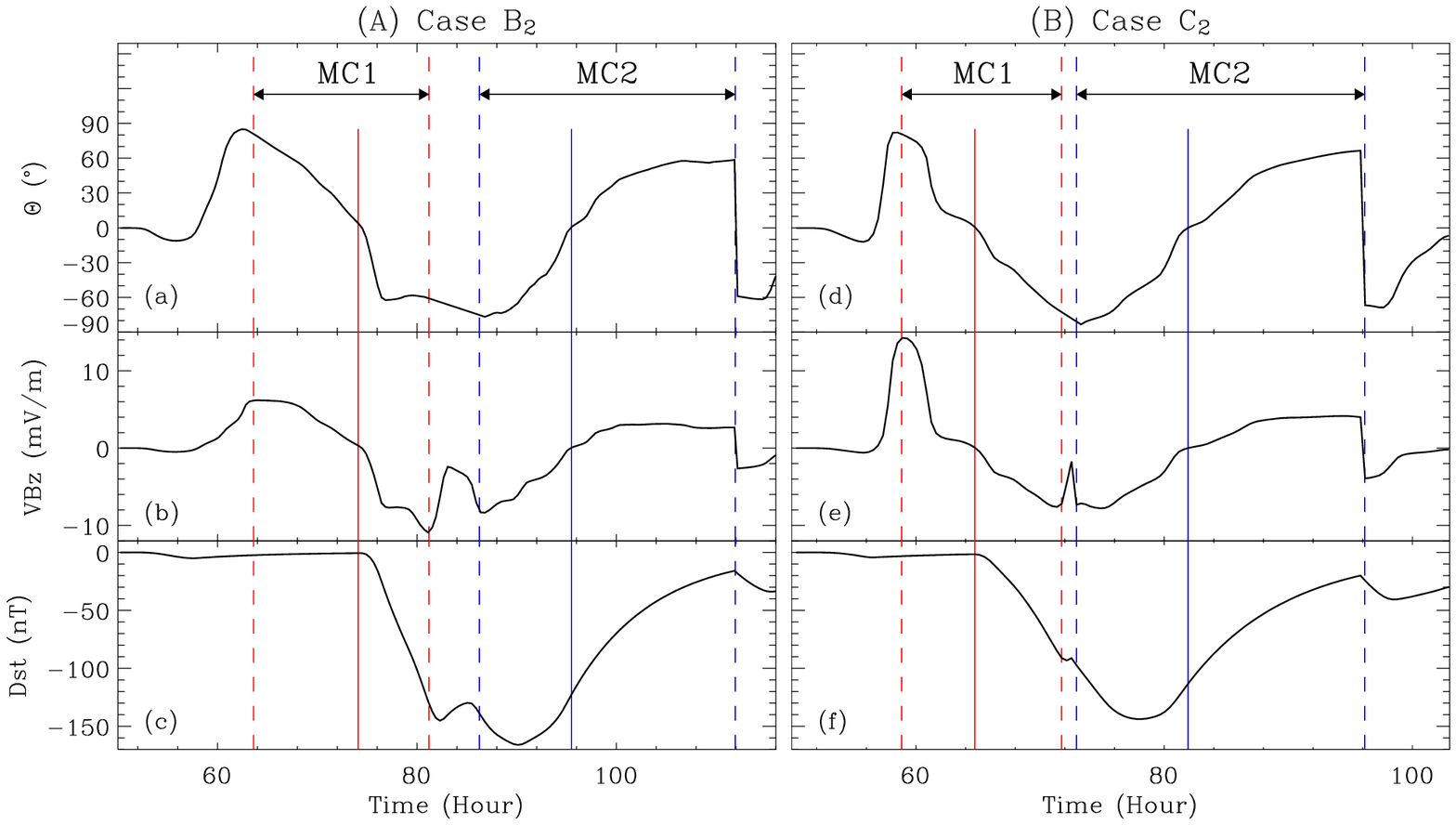}
\caption{} \label{Satellite.BC}
\end{figure}

There are various combination modes to form a double-MC structure
on basis of each sub-cloud helicity signature \citep{Wang2002},
one of which possessing the strongest geoeffectiveness is positive
helicity for preceding MC1 ($H_{mc1}=1$) and negative helicity for
following MC2 ($H_{mc2}=-1$) \citep{Wang2004}. According to this
scenario \citep{Wang2004}, simulation cases B$_2$ and C$_2$ are
run simply by reversing MC2 helicity in Cases B$_1$ and C$_1$,
respectively. The in-situ record of passage of multi-cloud event
at L1 is shown in Figure \ref{Satellite.BC}, with Columns (A) and
(B) corresponding to Cases B$_2$ and C$_2$, respectively. In
contrast to Figures \ref{Satellite-B} and \ref{Satellite-C}, the
elevation angle $\Theta$ of magnetic field vector within the
double-flux-rope structure in Figure \ref{Satellite.BC} is changed
from the north-south-north-south orientation to
north-south-south-north one. Though two $Dst$ dips exist in Groups
EID$_2$ and CID$_2$, close scrutiny reveals that (1) the recovery
phase of the first trivial $Dst$ dip is extremely short (3.3 and
0.9 hours in Cases B$_2$ and C$_2$, respectively); (2) the second
$Dst$ dip is low enough to describe the whole geoeffectiveness by
its local minimum, with $-166$ nT at 90 hours in Case B$_2$ and
$-144$ nT at 78 hours in Case C$_2$. Hence from the perspective of
continuous interval with southward magnetic field $B_s$, $Dst$
curve in Groups EID$_2$ and CID$_2$ can be considered as a one-dip
structure by ignoring the first trivial dip. The closer the
distance between two sources of IP geoeffective trigger, the
easier is the superposition of individual geoeffectiveness, the
greater is the resulting geomagnetic storm. This is confirmed by
contrast of Figures \ref{Satellite.BC}(c) and (f) with Figures
\ref{Satellite-B}(e) and \ref{Satellite-C}(e).

\section{Geoeffectiveness Studies}\label{Sec:Geoeffect}
\begin{figure}
   \includegraphics[width=0.99\textwidth]{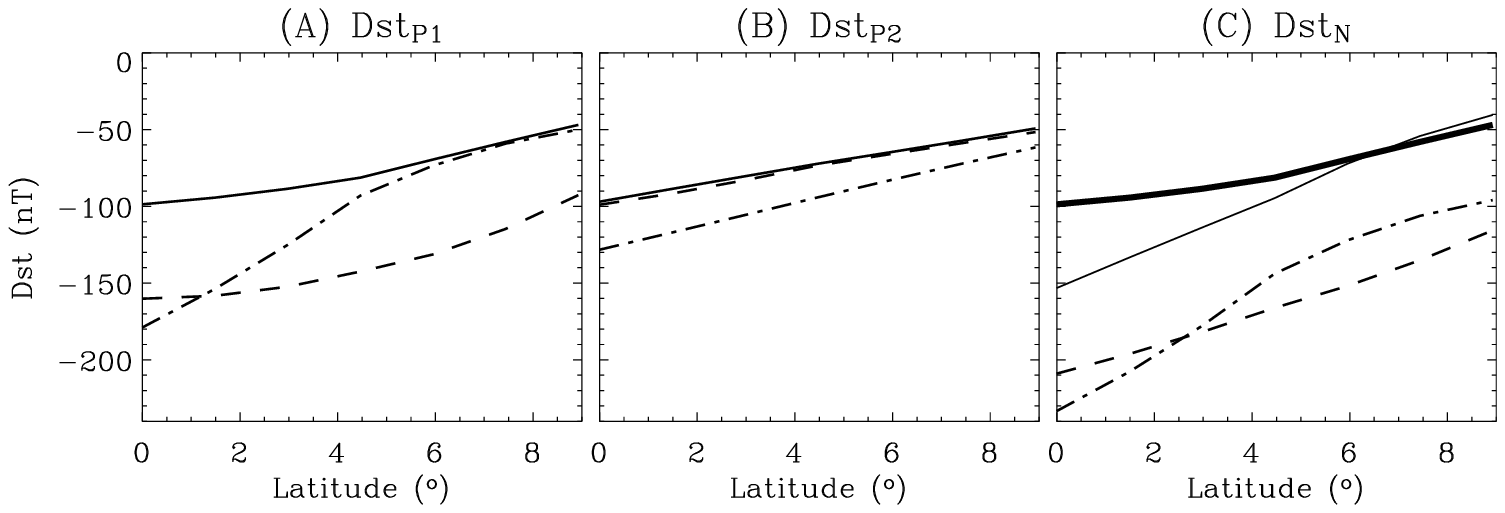}
\caption{} \label{Dst-Lat}
\end{figure}

Near-HCS latitudinal dependence of the $Dst$ index is plotted in
Figure \ref{Dst-Lat}, where Columns (A) $Dst_{P1}$ and (B)
$Dst_{P2}$ represent the first and the second $Dst$ dips in Cases
B$_1$ and C$_1$, meanwhile Column (C) $Dst_N$ depicts the single
$Dst$ dip in Cases B$_2$ and C$_2$. The dashed and dash-dotted
lines represent for Cases B$_1$ and C$_1$, respectively, in
Columns (A) and (B). And they represent for Cases B$_2$ and C$_2$,
respectively, in Column (C). As the MC2-driven shock continues to
propagate through the MC1 medium, $Dst_{P1}$ increases within Lat.
$> 1.3^\circ$ and decreases within Lat. $< 1.3^\circ$, found in
Figure \ref{Dst-Lat}(A). Meanwhile the distribution of $Dst_N$ in
Figure \ref{Dst-Lat}(C) is quite similar. The trend of decreased
$Dst$ near HCS is opposite to that in the case of MC-shock
interaction \citep[c.f. Figure 8 in][]{Xiong2006a}. The above
divergence is clarified by the absence of following MC body
pushing in MC-shock interaction \citep{Xiong2006a}. First, the
latitudinal extent of MC2 body is much narrower than that of
MC2-driven shock. Second, the coalescent boundary between MC1 and
MC2 body is further narrower, which covers latitude range between
$4.5^\circ$S and $4.5^\circ$N (Figure \ref{Case-C}(c)). Thus the
MC2 body pushing effect is strongest at the equator, within
confined latitudinal extent between $4.5^\circ$S and $4.5^\circ$N.
The near-HCS geoeffectiveness of $Dst_{P1}$ from Case B$_1$ to
Case C$_1$, $Dst_N$ from Case B$_2$ to Case C$_2$ is subsequently
aggravated. As a result, nonuniform latitudinal distribution of
$Dst_{P1}$ and $Dst_N$ is intensified. Besides, $Dst_{P2}$ is
nearly unaffected in Case B$_1$. However $Dst_{P2}$ in Case C$_1$
is obviously decreased, as a result from the compression of MC2
body interpreted in Section \ref{Sec:CaseC}. Hence the
geoeffectiveness of Multi-MC is indeed largely enhanced due to
interaction between sub-clouds, as compared with that in an
isolated MC event.

\begin{figure}
   \includegraphics[width=0.6\textwidth]{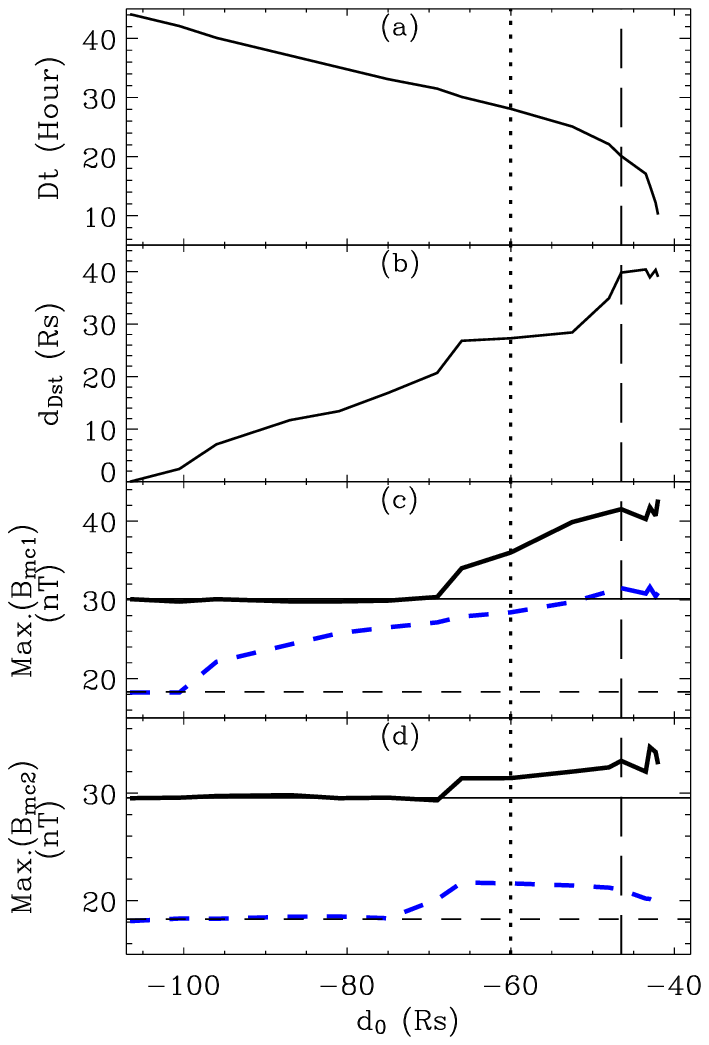}
\caption{} \label{Depth-DtB}
\end{figure}

In order to quantify the evolution process of Multi-Cloud, $d_0 =
r_{mc2} - r_{mc1}$, the distance between the cores of MC2 and MC1
on the occasion of MC1 head just reaching L1, is chosen as an
indicative parameter. $r_{mc1}$ and $r_{mc2}$ are the core
positions of MC1 and MC2 in radial direction respectively. The
reliance of several multi-cloud parameters on $d_0$ is further
explored in Figure \ref{Depth-DtB} by the integrated study of
Groups EID$_1$ and EID$_2$. The absolute value of $d_0$ is labeled
as $|d_0|$. As $Dt$, the emergence interval of MC1 and MC2,
decreases, $|d_0|$ firstly reduces from 107 to 53 $R_s$ at a
constant slope, then asymptotically approaches to 42 $R_s$ shortly
after MC2-driven shock emerges from MC1 body head (Figure
\ref{Depth-DtB}(a)). The penetration depth of MC2-driven shock in
MC1 medium, $d_{Dst}$, defined by the radial distance between
MC2-driven shock front and MC1 inner boundary along the equator,
is shown in Figure \ref{Depth-DtB}(b), which can be divided into
four stages according to the different behaviors: (1) a rapid
increase during $d_0< -66.3 R_s$, (2) an extremely slow increase
during $-66.3 R_s <d_0< -52.5 R_s$, (3) a fast re-increase during
$-52.5 R_s <d_0< -46.7 R_s$, (4) a very small oscillation around
the final limit value of 40 $R_s$ during $d_0> -46.7 R_s$. The
rapid increasing of $d_{Dst}$ in stages (1) and (3) is
straightforward due to continuous forward movement of shock front
in MC1 medium. As for stage (2) during which the shock front hits
MC1 core, though the shock front location relative to the MC1 body
is deeper and deeper at that time, the abrupt change of MC1 rear
boundary morphology from a V-shape to a straight line, mentioned
in Section \ref{Sec:CaseC}, greatly reduces the radial extent of
MC1 rear half, and hence significantly inhibits the increase of
the absolute value of penetration depth $d_{Dst}$. When the shock
front crosses the MC1 front boundary ($d_0 > - 46.7 R_s$), the
magnetic tension of the highly compressed MC1 body is drastically
accumulated. As a result, the nearer the distance between two
sub-MCs is (the shorter the $|d_0|$ is), the larger is the
resistance of MC1 elasticity against compression. The final
equilibrium is naturally manifested in the behavior of stage (4).
The early and sensitive response of Max.$(B_{mc1})$ at $d_0=-100
R_s$ is conspicuous along Lat. = $4.5^\circ$ in Figure
\ref{Depth-DtB}(c), because the initial interaction between MC1
body and MC2-driven shock happens around Lat. = $4.5^\circ$. The
swift enhancement of Max.$(B_{mc1})$ during $-68.9 R_s < d_0 <
-46.7 R_s$ along Lat. = $0^\circ$ is owing to the compression
concurrently exerted by the MC2-driven shock and MC2 body. Both
Max.$(B_{mc1})$ and Max.$(B_{mc2})$ reach a relatively stable
state when $|d_0|= 42 R_s$.

\begin{figure}[htbp]
\begin{center}
 \noindent
  \includegraphics[height=.99\textwidth,width=.95\textheight,keepaspectratio,angle=90]{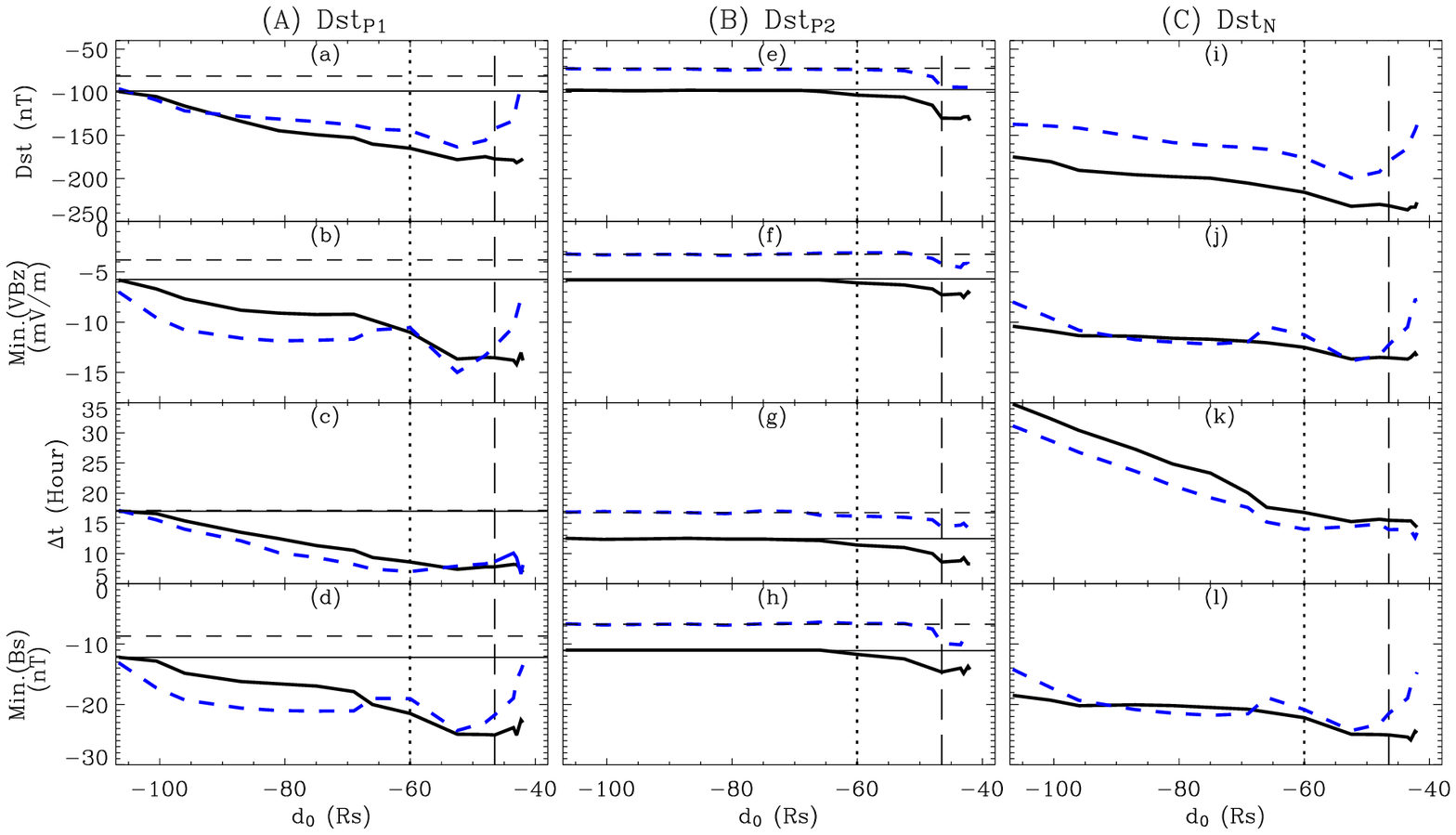}
\caption{} \label{depth}
\end{center}
\end{figure}

The variance of geoeffectiveness as a function of $d_0$ is
elucidated in Figure \ref{depth}. The analyses on $Dst_{P1}$,
$Dst_{P2}$, and $Dst_N$ are addressed one by one. First, when
$d_0< -60 R_s$, the behavior of all parameters in Figures
\ref{depth}(a)-(d) for $Dst_{P1}$ is pretty coincident with that
of our previous study for MC-shock interaction \citep[c.f. Figures
9(b)-(e) in][]{Xiong2006a}. The dynamics of MC1-MC2 merging at
that time is dominated by the interaction between MC2-driven shock
and MC1 body. Thus MC2-driven shock plays the similar role of the
incidental shock as addressed before \citep{Xiong2006a}, which
clarifies the above-mentioned coincidence. As $|d_0|$ is reduced
from $60R_s$ to $52.5R_s$, MC2 body directly collides with MC1
body. It leads to the decrease of $Dst$, Min.$(VB_z)$, and
Min.$(B_s)$ due to compression. Particularly, the decrease of
Min.$(VB_z)$ and Min.$(B_s)$ along Lat. = $4.5^\circ$S is very
drastic, because the change of MC1 field line morphology from a
V-shape to a straight line mentioned in Section \ref{Sec:CaseC}
leads to the southward rotation of magnetic field within MC1 rear
half along Lat. = $4.5^\circ$S. This additional rotation effect
further strengthens $B_s$ along Lat. = $4.5^\circ$S. When $|d_0|$
continues to decrease to be less than $52.5 R_s$, significant
difference of geoeffectiveness between Lat. = $0^\circ$ and
$4.5^\circ$S occurs. Along Lat. = $4.5^\circ$S, the rapid recovery
of Min.$(B_s)$ from $-24.5$ to $-13.5$ nT, and Min.$(VB_z)$ from
$-15$ to $-8$ mV/m, leads to the subdued $Dst_{P1}$ from its
minimum $-165$ to $-100$ nT. Contrarily, the geoeffectiveness
along Lat. = $0^\circ$ remain unchanged (Figures
\ref{depth}(a)-(d)). Namely, the aggravated geoeffectiveness along
the equator is the same with $Dst_{P1}=-180$ nT, provided that
$|d_0|$ is smaller than a certain threshold of 52.5 {$R_s$}. This
highly nonuniform latitudinal distribution of $Dst_{P1}$ is owing
to the limited latitudinal range ($4.5^\circ$S $\sim 4.5^\circ$N)
of pushing effect of MC2 body. When the shock ultimately
penetrates MC1 body, the persistent pushing of following MC2 body
within $4.5^\circ$S $\sim 4.5^\circ$N can prevent the previously
compressed magnetic field lines of MC1 body from being relaxed. So
$Dst_{P1}$ along Lat. $=0^\circ$ is nearly constant for $|d_0|<
52.5R_s$. As for $Dst_{P1}$ along Lat. $=4.5^\circ$S, it increases
as a result of relaxation of magnetic tension without MC2 body
pushing. Second, the variance of $Dst_{P2}$ (geoeffectiveness of
sub-MC2) only happens between $d_0 = -68 R_s$ and $-46.7 R_s$,
during which the MC2 body compression due to the blocking of MC1
body takes effect. Before the involving of MC2 body into
interaction ($d_0<-68R_s$), or after the completion of Multi-MC's
drastic evolution stage ($d_0>-46.7R_s$), $Dst_{P2}$ is unchanged.
By comparison $Dst_{P2}$ with $Dst_{P1}$, one can see that the MC1
undergoes the greater compression than the MC2. Third, the
behavior of $Dst_N$ (Figures \ref{depth}(i)-(l)) is quite similar
to that of $Dst_{P1}$ (Figures \ref{depth}(a)-(d)) due to similar
reasons mentioned above. The minimum $Dst$ in Figures
\ref{depth}(a), (e), and (i) is $-180$, $-130$, and $-235$ nT,
respectively. The greatest geoeffectiveness of $Dst_N$ directly
results from the longest $\Delta t$ (Figure \ref{depth}(k)).
Therefore, the geoeffective parameters of every sub-MC are
dramatically changed in contrast with those of the corresponding
isolated MC during the merging process. For the IP compound
structure formed by multiple ICMEs, the geoeffectiveness is
jointly determined by two factors: the parameters of the
individual ICMEs themselves, and the interaction process between
these ICMEs. This is substantiated by the observation data
analyses \citep{Wang2002,Wang2003a,Xue2005,Farrugia2006,Zhang2007}
and our quantitative investigation of numerical simulation of this
study.

\begin{figure}[htbp]
   \includegraphics[width=0.98\textwidth]{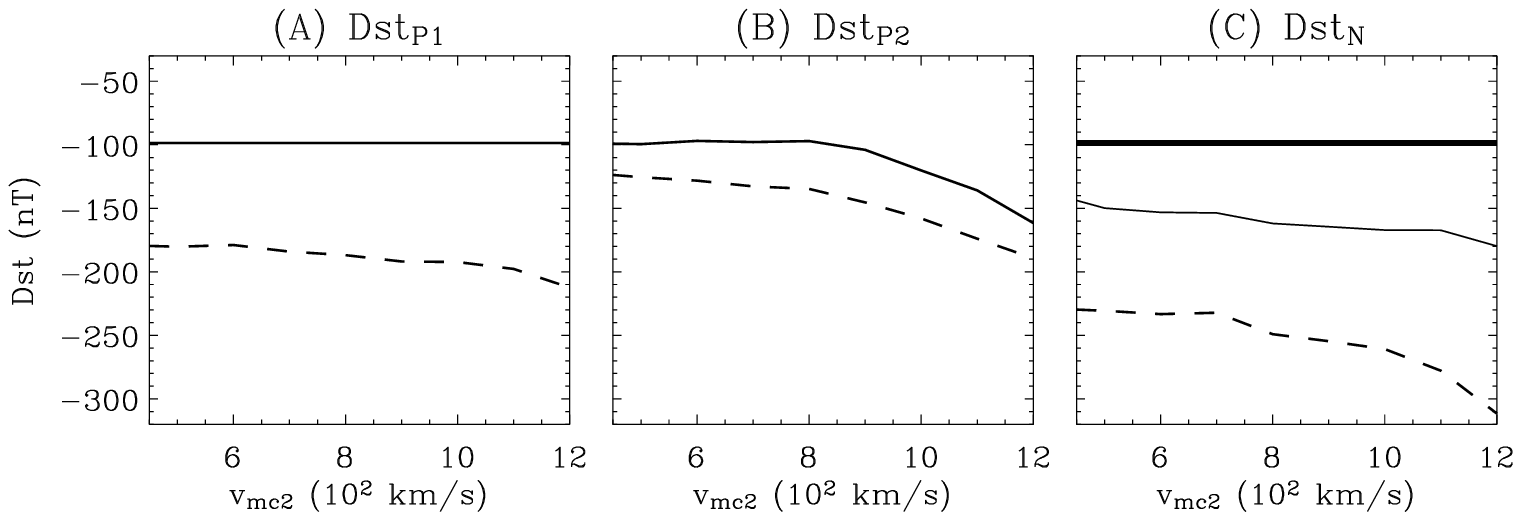}
\caption{} \label{mxiong-CID}
\end{figure}

The Multi-Cloud geoeffectiveness depends on not only the MC1-MC2
eruption interval, but also collision intensity. Obviously, an MC1
overtaken by an MC2 with various initial speeds may result in
different geoeffectiveness. From the Figure \ref{depth} concerning
Groups EID$_1$ and EID$_2$, two basic results are obtained: (1)
The maximum geoeffectiveness occurs at Lat. = $0^\circ$ for the
same propagation direction of MC1 and MC2 along the equator; (2)
The final $Dst$ at Lat. = $0^\circ$ is nearly constant, provided
the accompanying $|d_0|$ is sufficiently small ($|d_0| \leq 46.7
R_s$), or the initial MC1-MC2 eruption interval is sufficiently
short ($Dt \leq 20 $ hours). With $t_{mc2}$ designated to be 12.2
hours, the reliance of geoeffectiveness along the equator on
collision degree is further explored in Figure \ref{mxiong-CID} by
parametric study of variable $v_{mc2}$. The larger the value of
$v_{mc2}$ is, the greater is the collision degree that the
Multi-MC may suffer from. $Dst_{P1}$ only decreases a bit from
$-180$ to $-210$ nT within such a wide spectrum of $v_{mc2}$ from
450 to 1200 km/s. The geoeffectiveness enhancement of Multi-MC is
ascribed to compression between the sub-MCs. When the MC1
compression has already approached to saturation, the effect to
increase MC1 geoeffectiveness by having MC1 impinged by a highly
fast MC2 is extremely limited. It is more and more difficult to
quench the dramatically accumulated magnetic elasticity of MC1
body, as MC1 undergoes the greater and greater compression. The
impact of the high-speed MC2 body is largely offset by the buffer
action of magnetic tension of the MC1 body. As for $Dst_{P2}$, the
increase of $v_{mc2}$ has a direct influence. However, $Dst_{P2}$
deducted by the $Dst$ of the corresponding individual MC2 event is
roughly constant, which can be seen from Figure
\ref{mxiong-CID}(b). Namely $Dst_{P2}$ decreases from $-125$ to
$-190$ nT, as $v_{mc2}$ increases from 450 to 1200 km/s, chiefly
ascribed to the increase of geoeffectiveness of the corresponding
individual MC2 event itself, but not MC1-MC2 interaction.
Excluding the geoeffectiveness increase of individual MC2 event,
$Dst_N$ still decreases for $v_{mc2}>$ 1000 km/s in Figure
\ref{mxiong-CID}(c), because interaction takes obvious effect
herein. The geoeffectiveness variance can be elucidated from the
perspective of dynamic response of sub-MCs. The double-MC
interacting region is within MC1 rear part and MC2 front part,
where the direct compression occurs. So the factor of MC1-MC2
interaction for geomagnetic storm enhancement is strongest for
$Dst_N$, weakest for $Dst_{P2}$. In conclusion, two points can be
drawn from Figures \ref{depth} and \ref{mxiong-CID}: (1) The
significant geoeffectiveness variance accompanies the different
evolution stages; (2) Once a Multi-MC completes its evolution
process before its arrival at 1 AU, the collision intensity
between sub-MCs merely modulates the final geoeffectiveness a bit.
The innate magnetic elasticity can buffer the reciprocal collision
between sub-MCs against each other. When every sub-MC becomes
stiffer and stiffer, the compression reaches its asymptotic
degree, and the geoeffectiveness enhancement becomes less and less
obvious. Therefore, with respect to Multi-MC geoeffectiveness, the
evolution stage is a dominant factor, whereas the collision
intensity is a subordinate one.

Additionally, the dependence of geoeffectiveness of an individual
MC on the eruption speed $v_{mc}$ is also revealed from the
isolated MC2 events from Figure \ref{mxiong-CID}. If $B_s$ region
in MC medium is located in its anterior half (Group IM$_2$), $Dst$
steadily decreases as $v_{mc}$ increases, as seen by the thin
solid line in Figure \ref{mxiong-CID}(C); Contrarily, if $B_s$
region is to be in the rear half of MC (Group IM$_1$), $Dst$ only
decreases on the condition of $v_{mc}>$ 800 km/s, as seen by the
solid line in Figure \ref{mxiong-CID}(B). The increase of $v_{mc}$
leads to a more violent interaction of individual MC body with the
ambient solar wind ahead. As a result, MC core, initially located
at the geometry center of MC boundary, will be gradually shifted
to MC anterior boundary. MC anterior half is preferential
compressed, because MC-ambient flow interaction originates from MC
front boundary. The compression exists in MC rear half, only when
the whole cross section area of MC body is significantly
contracted on the condition of very fast speed $v_{mc}$. This is
why $Dst$ for Group IM$_1$ remains a constant of $-100$ nT within
$v_{mc}$ = 450 $\sim$ 800 km/s.

\section{Compressibility Analyses} \label{Sec:Compressibility}
The idea that the compression is an efficient mechanism to enhance
the geoeffectiveness of the pre-existing $B_s$ event has been
proved in data analyses \citep{Wang2003c}. Compression effect is
virtually responsible for the geoeffective property of Multi-MC.
So it is very meaningful to analyze the maximum compression degree
for a Multi-MC.

\begin{figure}[htbp]
   \includegraphics[width=0.99\textwidth]{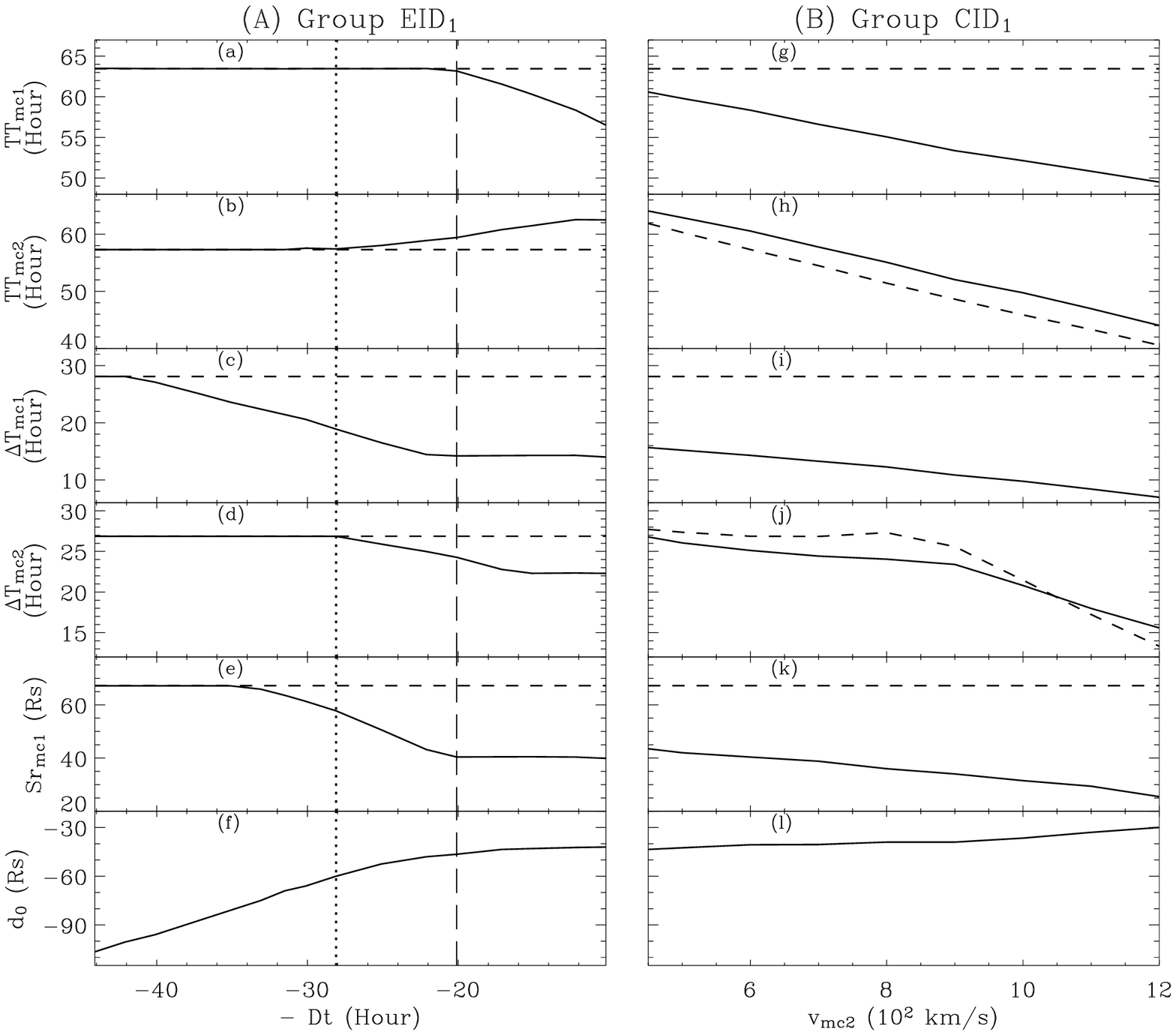}
\caption{} \label{wym}
\end{figure}

The Multi-MC characteristics can be inferred from several
parameters of near-Earth measurements, depicted by Figure
\ref{wym}. The interchange of momentum between the preceding slow
cloud MC1 and following fast cloud MC2 leads to MC1 acceleration
and MC2 deceleration, which influences, more or less, the
Sun-Earth transient time, $TT_{mc1}$ and $TT_{mc2}$ for MC1 and
MC2, respectively. The shortening of $TT_{mc1}$ begins at $Dt =
21$ hours as seen from Figure \ref{wym}(a), meanwhile the
lengthening of $TT_{mc2}$ begins at $Dt = 28$ hours, seen from
Figure \ref{wym}(b). The MC1 acceleration is very obvious, as the
larger $v_{mc2}$ is, the smaller is $TT_{mc1}$ (Figure
\ref{wym}(g)). Contrarily, the MC2 slowdown is independent of
$v_{mc2}$, as $TT_{mc2}$ in coupled events deviates from that in
the corresponding isolated events by a constant (Figure
\ref{wym}(h)). The effect of $TT_{mc1}$ decrease is much greater
than that of $TT_{mc2}$ increase for Multi-MC cases. Since the
transporting time of an ICME may be modified if it interacts with
others during its IP propagation, some empirical formulas of
transporting time on basis of observations of one single ejecta
event \citep{Gopalswamy2000,Gopalswamy2001a} can not be directly
applied to the ICME-ICME interaction cases
\citep{Farrugia2004,Wang2005b,Xiong2005}. Coupling between ICMEs
occupies a large fraction for the causes of great geomagnetic
storms \citep{Xue2005,Zhang2007}, the Multi-MC should be paid
special attention for space weather predicting. Thus the numerical
simulation based on physics models is very useful to forecast the
arrival time of the interacting ICMEs. The duration of sub-MC
passage at L1, $\Delta T_{mc}$, is a distinct reflection of
compression effect. $\Delta T_{mc1}$ exists a lower limit, as
shown in Figure \ref{wym}(c), so does $\Delta T_{mc2}$ in Figure
\ref{wym}(d). When the Multi-MC experiences the sufficient
evolution for $Dt < 24$ hours, the reduction of $\Delta T_{mc1}$
and $\Delta T_{mc2}$ is 14 and 4.5 hours respectively, in contrast
with the corresponding isolated sub-MC cases. As $v_{mc2}$
increases, both $\Delta T_{mc1}$ and $\Delta T_{mc2}$
monotonically decrease. However, the solid and dashed lines,
representing the Multi-MC and corresponding isolated MC events in
Figure \ref{wym}(j), intersect at $v_{mc2}=$ 1040 km/s. $\Delta
T_{mc2}$ in MC1-MC2 interaction is determined by two factors: (1)
the compression of MC2 radial extent resulting from collision; (2)
the slowdown of MC2 body as a result of momentum transfer from MC2
to MC1 body. The first factor, tending to shorten $\Delta
T_{mc2}$, dominates the cases for $v_{mc2}<$ 1040 km/s; the second
factor, trending to lengthen $\Delta T_{mc2}$, dominates the cases
for $v_{mc2}>$ 1040 km/s. Besides, the near-Earth radial span of
MC1 body $Sr_{mc1}$ in Figures \ref{wym}(e) and (k) has the
similar variance trend as $\Delta T_{mc1}$ in Figures \ref{wym}(c)
and (i). It again proves that the compression has saturation
effect for MC1 body. The $Sr_{mc1}$ of 67 $R_s$ in an individual
case can be compressed to 40 $R_s$ at most by $Dt$ reduction
(Figure \ref{wym}(e)). $Sr_{mc1}$ decreases very slowly from 43
$R_s$ at $v_{mc2}=$ 450 km/s to 25 $R_s$ at $v_{mc2}=$ 1200 km/s
(Figure \ref{wym}(k)). Moreover, the overall compression degree
for a Multi-MC is well described by $d_0$, the distance between
the core of the following MC2 and preceding MC1 on the occasion of
MC1 head just reaching L1. One can see that $d_0$ variance is
associated with Multi-MC evolution stages (Figure \ref{wym}(f)).
The swiftly reducing trend of $|d_0|$ at the beginning is suddenly
stopped at $Dt = $ 25 hours. $|d_0|$ reaches its lower limit of 42
$R_s$ at $Dt = $ 17 hours, and maintains a horizontal slope
afterwards. When the inherent magnetic tension rivals the external
compression for force balance, each sub-MC behaves like a rigid
body with a little elasticity. $|d_0|$ is only reduced from 43 to
30 $R_s$ over such a wide $v_{mc2}$ range from 450 to 1200 km/s
(Figure \ref{wym}(l)).

The compression due to interaction is primarily responsible for
geoeffectiveness enhancement, once two MCs form a Multi-MC.
Assuming nonexistence of magnetic field in the IP medium and all
ejecta, the preceding ejecta may be exorbitantly compressed to an
unbelievably small scale by the following ejecta
\citep{Gonzalez-Esparza2004,Gonzalez-Esparza2005}. Obviously the
compressibility on basis of hydrodynamic nature
\citep{Gonzalez-Esparza2004,Gonzalez-Esparza2005} is overestimated
due to ignoring of magnetic elasticity. The larger the compression
is, the stiffer is every sub-MC body. Hence a cutoff compression
degree exists because of magnetic tension. Besides, if the
helicity of MC1 is consistent with that of MC2, a electric current
sheet occurs between the adjoining boundary of MC1 and MC2 due to
magnetic field direction reversion. The electric current intensity
synchronously increases with the Multi-MC compression. If magnetic
reconnection happens there, the MC1-MC2 collision effect would be
weakened. As a result, the outermost part of magnetic field lines
of each sub-MC would be reconnected together \citep{Wang2005b}.
Particularly in the condition of large speed difference between
MC1 and MC2, both MCs may be merged into one new magnetic flux
rope by the driven magnetic reconnection
\citep{Odstrcil2003,Schmidt2004,Wang2005a}. The magnetic
reconnection reduces Multi-MC's cutoff compression degree. If
magnetic reconnection is introduced into Groups EID$_1$ and
CID$_1$ of Table \ref{Tab2}, the Multi-MC geoeffectiveness would
become weakening due to the subdued compression and south magnetic
component annihilation. However, magnetic diffusion in the IP
space should be very small, magnetic reconnection may sightly
modulate, but not significantly distort the dynamics and
geoeffectiveness of Multi-MC in the framework of ideal MHD
process. So the CME-CME cannibalization, firstly observed in the
inner corona by the SOHO/Lasco \citep{Gopalswamy2001b}, later
proved to be caused by magnetic reconnection \citep{Wang2005a},
may not occur in the IP space \citep{Wang2005b}.

\section{Conclusions and Summary}\label{Sec:Conclusion}
In order to better understand the nature of IP Multi-MC structure,
the interaction between two IP MCs (MC1 and MC2), and the ensuing
geoeffectiveness are explored under a very simplified and
specialized circumstance by a 2.5-dimensional ideal MHD numerical
model. This work is a continuation to our recent studies of
MC-shock interaction \citep{Xiong2006a,Xiong2006b} by replacing a
following incidental strong shock with a following fast MC. Via
analyses of a comprehensive integration of many simulation cases
under various conditions, it is found that the magnetic
elasticity, magnetic helicity of each MC, and compression between
each other are the overriding physical factors in the formation,
propagation, evolution, and resulting geoeffectiveness of IP
Multi-MC.

First, the dynamical response of MCs colliding is studied. The
coupling of two MCs could be considered as the comprehensive
interaction between two systems, each comprising of an MC body and
its driven shock. Because the following MC2 is faster than the
preceding MC1, the MC2-driven shock and MC2 body successively
impact the rear boundary of MC1 body. As a result, the morphology
of magnetic field lines at MC1's rear part is consequently changed
from its initial rough semi-circle to a V-shape, and then to a
flat line. As swept by the marching MC2-driven shock front, the
local magnetic field lines in MC1 medium just downstream of
MC2-driven shock front would be compressed and rotated. The
pushing of MC2 body prevents the previously compressed magnetic
field in MC1 medium from being restored, after the passage of
MC2-driven shock front. MC1 body undergoes the most violent
compression from the ambient solar wind ahead, continuous
penetration of MC2-driven shock through MC1 body, and persistent
pushing of MC2 body at MC1 tail boundary, which leads to a
significant shrinking of MC1's cross section. Contrarily, the
blocking of MC1 body also results in the change of MC2 boundary
from a radial-extent-elongated ellipse to an
angular-extent-elongated one. The Momentum is continuously
transferred from sub-MC2 to sub-MC1, until the radial profile of
Multi-MC speed is monotonically decreasing with the maximum value
at MC1-driven sheath. When MC2-driven shock is merged with
MC1-driven shock into a stronger compound shock, Multi-MC
completes its ultimate evolutionary stage, and hence moves forward
as a relatively stable entity.

Second, the geoeffectiveness of MCs coupling is explored. The
interaction of MC1 and MC2 in the IP space results in the
superposing of their geoeffectiveness. The two-MC event is
associated with a two-step geomagnetic storm, as indicated by two
$Dst$ dips. Particularly, if $B_s$ region in a Multi-MC is located
at MC1 rear half and MC2 anterior half, the Multi-MC excites the
greatest geomagnetic storm among all combinations of each sub-MC
helicity, and two $Dst$ dips can be nearly reduced to a single
$Dst$ dip due to ignoring of the very short recovery phase of the
first $Dst$ dip. The geoeffectiveness of each individual MC is
largely enhanced as a result of MC1-MC2 interaction. Moreover,
because latitudinal extent of MC body is much narrower than that
of its driven shock, the effect of MC2 body pushing upon MC1 body
is limited within a very narrow latitudinal band centered at the
heliospheric equator. Outside this latitudinal band,
geoeffectiveness is initially enhanced and then recovered, as the
emergence interval of two MCs becomes shorter and shorter;
meanwhile the geoeffectiveness is firstly aggravated and then
maintains constant inside this band. Obviously, the nonuniform
latitudinal distribution of geoeffectiveness is further
intensified by MC2 body pushing. Moreover, With respect to
Multi-MC geoeffectiveness, the evolution stage is a dominant
factor, whereas the collision intensity is a subordinate one.

Third, Multi-MC's compressibility associated with magnetic
elasticity is analyzed. Both compression degree and evolutionary
stage of a Multi-MC could be quantitatively described by $|d_0|$,
the absolute distance between MC1 and MC2 core on the occasion of
MC1 head just reaching L1. The shorter the $|d_0|$ is, the greater
is Multi-MC's compressibility. Magnetic field lines of MC1 body
initially appears to be too frail to resist the collision in the
face of the overtaking high-speed MC2, so $|d_0|$ is steadily
reduced. As the evolution of Multi-MC proceeds, the MC1 body
suffers from larger and larger compression, and its original
vulnerable magnetic elasticity becomes stiffer and stiffer. When
the accumulated inherent magnetic elasticity in the highly shrunk
MC1 body can counteract the external compression, the previous
continuously reducing $|d_0|$ drastically approximates to an
asymptotic limit. Magnetic elasticity not only buffers the
collision between MCs, but also leads to a cutoff compression
degree of Multi-MC. Moreover, the collision of MC2 with a very
wide speed spectrum upon MC1 has a little influence to enhance the
cutoff compressibility. However, if magnetic reconnection occurs
within the interacting region of Multi-MC, the cutoff
compressibility would be expected to decrease a bit.

Overall, the Multi-MC is of great concern for space weather
community. The geoeffectiveness enhancement of coupling of
multiple MCs is virtually ascribed to compression in the Multi-MC.
The maximum compressibility of Multi-MC is mainly decided by its
inherent magnetic elasticity.

\begin{acknowledgments}
This work was supported by the National Key Basic Research Special
Foundation of China (2006CB806304), the Chinese Academy of
Sciences Grant No. KZCX3-SW-144, the National Natural Science
Foundation of China (40336052, 40404014, 40525014 and 40574063),
and the Chinese Academy of Sciences (startup fund). S. T. Wu was
supported by an NSF grant (ATM03-16115).
\end{acknowledgments}

\bibliography{Xiong}



%
%
%

\section*{Figure Captions}
\begin{description}
\item[Figure 1] The evolution of MC2 overtaking MC1 for Case
B$_1$, with (a)-(c) magnetic field magnitude $B$, (d)-(f) radial flow speed $v_r$, and (g)-(i) radial
characteristic speed of fast mode $c_f$. Attached below each image are two additional radial profiles
along Lat.$=0^\circ$ and $4.5^\circ$S. Note that radial profile of $B$ is plotted by subtracting the
initial ambient value $B|_{t=0}$. The white solid line in each image denotes the MC boundary. Solid
and dashed lines at each profile denote MC core and boundary. Only part of domain is adaptively
plotted to highlight Multi-MC.

\item[Figure 2] In-situ hypothetical observation along Lat. =
$4.5^\circ$S for Case B$_1$. Stacked from top to bottom are (a)
magnetic field magnitude $B$, (b) elevation of magnetic field
$\Theta$, (c) radial flow speed $v_r$, (d) derived dawn-dusk
electric field $VB_z$, and (e) $Dst$ index.  Solid and dashed
delimiting lines denote MC center and boundary.

\item[Figure 3] The evolution of MC2 overtaking MC1 for Case C$_1$
with radial flow speed $v_r$.

\item[Figure 4] In-situ hypothetical observation along Lat. =
$4.5^\circ$S for Case C$_1$.

\item[Figure 5] The time dependence of MC parameters: (a) radial
distance of MC core $r_m$, (b) MC radial span $Sr$, (c) MC angular
span $S\theta$, and (d) MC cross section area $A$. The thick
dashed and solid lines denoted the preceding MC1 and following MC2
in Multi-MC Case C$_1$, superimposed with thin lines for
corresponding individual MC cases for comparison. Three vertical
delimiting lines (dotted, dashed and dotted) from left to right
correspond to the occasion of MC2-driven shock encountering MC1
body tail, MC2 body hitting MC1 body tail, and MC2-driven shock
reaching MC1 body head, respectively.

\item[Figure 6] In-situ hypothetical observation along Lat. =
$4.5^\circ$S for (A) Case B$_2$ and (B) Case C$_2$. Cases B$_2$ and C$_2$ differ from their respective
companion Cases B$_1$ and C$_1$ by the opposite MC2 magnetic helicities.

\item[Figure 7] The comparison of latitudinal distribution of
$Dst$ index among the Multi-MC Cases B$_1$, C$_1$, B$_2$, and
C$_2$. Double $Dst$ dips in Cases B$_1$ and C$_1$ with positive
magnetic helicities in MC2 are shown in (A) $Dst_{P1}$ and (B)
$Dst_{P2}$, as well as a single $Dst$ dip in Cases B$_2$ and C$_2$
with negative helicity in MC2 shown in (C) $Dst_N$. Dashed and
dash-dotted lines in (A) and (B) correspond to Cases B$_1$ and
C$_1$ respectively; dashed and dash-dotted lines in (C) correspond
to Cases B$_2$ and C$_2$ respectively. The isolated events
corresponding to MC1 and MC2 for Case B$_1$ are denoted as solid
lines in (A) and (B), those for Case B$_2$ as solid thick and thin
lines in (C).

\item[Figure 8] The $d_0$-dependent parameter variances at L1 in
Group EID$_1$: (a) $Dt$, time interval of MCs
($Dt=t_{mc2}-t_{mc1}$, $t_{mc1}=0$ hour); (b) $d_{Dst}$,
penetration depth of MC2-driven shock in MC1 medium; (c)
Max.($B_{mc1}$), the maximum of magnetic field strength in MC1;
and (d) Max.($B_{mc2}$), the maximum of magnetic field strength in
MC2. Here $d_0$ refers to the distance between MC2 core $r_{mc2}$
and MC1 core $r_{mc1}$ on the occasion of MC1 head just reaching
L1, namely $d_0 = r_{mc2} - r_{mc1}$. The vertical delimiting
dotted and dashed lines denote the occasions of MC2-driven shock
just hitting MC1 core and head at L1. In (c) and (d), the thick
solid and dashed lines denote observations along Lat. = $0^\circ$
and $4.5^\circ$S, while the thin horizontal ones represent the
values of corresponding isolated MC events.

\item[Figure 9] The parameter variances of Multi-MC
geoeffectiveness as a function of $d_0$: (a, e, i) $Dst$ index;
(b, f, j) $~\mbox{Min.} (VB_z)$, the minimum of dawn-dusk electric
field $VB_z$; (c, g, k) $\Delta t$, the interval between the
commencement of $VB_z<-0.5~\mbox{ mV/m}$ and the corresponding
minimum $Dst$; and (d, h, l) $~\mbox{Min.}(Bs)$, the minimum of
southward magnetic component. Solid and dashed lines correspond to
observations along $~\mbox{Lat.= } 0^\circ$ and $4.5^\circ$S
respectively. The double $Dst$ dips in Group EID$_1$ are shown by
Columns (A) $Dst_{P1}$ and (B) $Dst_{P2}$, and the single $Dst$
dip in Group EID$_2$ by Column (C) $Dst_N$. The horizontal solid
and dashed lines denote observations of the isolated events,
corresponding to Group EID$_1$, at Lat. = $0^\circ$ and
4.5$^\circ$S respectively, with MC1 in Column (A) and MC2 in
Column (B).

\item[Figure 10] The reliance of $Dst$ in Multi-MC cases on
initial speed of following MC2. Double $Dst$ dips in Group CID$_1$
are shown as dashed lines by (A) $Dst_{P1}$ and (B) $Dst_{P2}$,
while a single $Dst$ dip in Group CID$_2$ by (C) $Dst_{N}$.
Decoupled MC1 and MC2 events in Group CID$_1$ are plotted as solid
lines in (A) and (B), those in Group CID$_2$ as thick and thin
solid lines in (C). The curves of single MC2 event in (B) and (C)
are non-horizontal due to $v_{mc2}$ variance.

\item[Figure 11] The dependence of Multi-MC characteristic
parameters at L1 on MC2-MC1 eruption interval $-Dt$ ($-Dt = -1
\cdot Dt= t_{mc1} - t_{mc2}$) and MC2 speed $v_{mc2}$ is shown by
Columns (A) Group EID$_1$ and (B) Group CID$_1$. $TT_{mc1}$, the
Sun-Earth transient time of MC1, is shown in (a) and (g); $\Delta
T_{mc1}$, MC1 event duration at L1, in (c) and (i); $Sr_{mc1}$,
MC1 radial span, in (e) and (k); and $d_0$, the distance between
MC2 and MC1 core, in (f) and (l). (a), (c), (g), (i) and (b), (d),
(h), (j) are the counterparts for MC1 and MC2 respectively. (e),
(f), (k), and (l) refer to the occasion when MC1 head just reaches
L1. Dashed lines in all panels except (f) and (l) represent the
corresponding isolated MC events for comparison. The vertical
dotted and dashed lines in Column (A) denote the cases of
MC2-driven shock just hitting MC1 core and head respectively.
\end{description}

\newpage
\begin{table}
\vspace*{2cm} %
\caption{Assortment of simulation cases of
individual MC}\label{Tab1}
\begin{tabular}{|l|l|l|l|}
\hline
Group & Case & $v_{mc}$ & Comment\\[0pt]
&& ($10^2$ km/s) &\\[0pt]
\hline
IM$_1$ & b$_1$, c$_1$, d$_1$, & 4, 6, 5, & Individual MC \\[0pt]
& e$_1$, f$_1$, g$_1$, & 7, 8, 9, & ($H_{mc}=1$) \\[0pt]
& h$_1$, i$_1$, j$_1$ & 10, 11, 12 &\\[0pt]
\hline
IM$_2$ & b$_2$, c$_2$, d$_2$, & 4, 6, 5, & Individual MC \\[0pt]
& e$_2$, f$_2$, g$_2$, & 7, 8, 9, & ($H_{mc}=-1$) \\[0pt]
& h$_2$, i$_2$, j$_2$ & 10, 11, 12 &\\[0pt]
\hline  
\end{tabular}
\end{table}

\newpage
\begin{table} 
\vspace*{2cm}%
\caption{Assortment of simulation cases of Multi-MC. Note that
$v_{mc1} = 400$ km/s, $t_{mc1}= 0$ hour for all 48
cases.}\label{Tab2}
\begin{tabular}{|l|l|l|l|l|}
\hline
Group & Case & $v_{mc2}$ & $t_{mc2}$ & Comment\\[0pt]
&&($10^2$ km/s) & (hour) &\\[0pt]
\hline  
EID$_1$ & B$_1$, C$_1$, D$_1$, E$_1$, & 6 & 30.1, 12.2,
44.1, 42.1, & Eruption Interval\\[0pt]
& F$_1$, G$_1$, H$_1$, I$_1$, && 40.1, 37.1, 35.1,
33.1, & Dependence\\[0pt]
& J$_1$, K$_1$, L$_1$, M$_1$, && 31.5, 28.1,
25.1, 22.1, & ($H_{mc1}=1$, $H_{mc2}=1$)\\[0pt]
& N$_1$, O$_1$, P$_1$, Q$_1$ && 20.1, 17.1,
15.1, 10.2, & \\[0pt]
\hline EID$_2$ & B$_2$, C$_2$, D$_2$, E$_2$, & 6 & 30.1, 12.2,
44.1, 42.1, & Eruption Interval \\[0pt]
& F$_2$, G$_2$, H$_2$, I$_2$, && 40.1, 37.1, 35.1,
33.1, & Dependence \\[0pt]
& J$_2$, K$_2$, L$_2$, M$_2$, && 31.5, 28.1,
25.1, 22.1, & ($H_{mc1}=1$, $H_{mc2}=-1$)\\[0pt]
& N$_2$, O$_2$, P$_2$, Q$_2$ && 20.1, 17.1,
15.1, 10.2, & \\[0pt]
\hline  
CID$_1$ & R$_1$, S$_1$, C$_1$, T$_1$, & 4.5, 5, 6, 7, & 12.2 & Collision Intensity \\[0pt]
& U$_1$, V$_1$, W$_1$, X$_1$, & 8, 9, 10, 11, && Dependence \\[0pt]
& Y$_1$ & 12 && ($H_{mc1}=1$, $H_{mc2}=1$)\\[0pt]
\hline
CID$_2$ & R$_2$, S$_2$, C$_2$, T$_2$, & 4.5, 5, 6, 7, & 12.2 & Collision Intensity \\[0pt]
& U$_2$, V$_2$, W$_2$, X$_2$, & 8, 9, 10, 11, && Dependence \\[0pt]
& Y$_2$ & 12 && ($H_{mc1}=1$, $H_{mc2}=-1$)\\[0pt]
\hline  
\end{tabular}
\end{table}

\clearpage
\newpage

\newpage

\newpage

\newpage

\newpage

\newpage

\newpage

\newpage

\newpage


\newpage

\newpage

\clearpage

\end{article}
\end{document}